\documentclass[11pt, a4paper]{article}
\pdfoutput=1 
\usepackage{jcappub}

\usepackage[T1]{fontenc} 


\usepackage[utf8]{inputenc}
\usepackage{subcaption}
\usepackage{setspace}
\usepackage{setspace}
\usepackage{graphicx,color}
\usepackage{epstopdf}
\usepackage{setspace}
\usepackage{makeidx} 
\usepackage{lmodern}

\usepackage[T1]{fontenc}
\usepackage[utf8]{inputenc}
\usepackage{color}
\usepackage{booktabs}
\usepackage{units}
\usepackage{amsmath}
\usepackage{amssymb}
\usepackage{esint}
\usepackage{braket}

\usepackage{caption}
\usepackage{subcaption}
\usepackage[export]{adjustbox}
\pdfpageheight\paperheight
\pdfpagewidth\paperwidth

\def\no{\nonumber}

\def\bea{\begin{eqnarray}}  
\def\eea{\end{eqnarray}}

\def\be{\begin{equation}}
\def\ee{\end{equation}}

\def\no{\nonumber}

\def\H0{H_{0}}

\def\Mp{M_{p_0}}
\def\Mpk{M_{p_k}}

\makeatletter

\newcommand{\Rmnum}[1]{\expandafter\@slowromancap\romannumeral #1@}
\makeatother

\usepackage{anysize}
\marginsize{2.5cm}{2cm}{1cm}{3.5cm}

\usepackage{natbib}
\title{\boldmath  Higgs inflation and quantum gravity: An exact renormalisation group approach}
\author{Ippocratis D. Saltas}
\emailAdd{isaltas@fc.ul.pt}
\affiliation{Institute of Astrophysics and Space Sciences, Faculty of Sciences, Campo Grande, PT1749-016 Lisboa, Portugal}
\abstract{We use the Wilsonian functional Renormalisation Group (RG) to study quantum corrections for the Higgs inflationary action including the effect of gravitons, and analyse the leading-order quantum gravitational corrections to the Higgs' quartic coupling, as well as its non--minimal coupling to gravity and Newton's constant, at the inflationary regime and beyond. 
We explain how within this framework the effect of Higgs and graviton loops can be sufficiently suppressed during inflation, and we also place a bound on the corresponding value of the infrared RG cut--off scale during inflation. Finally, we briefly discuss the potential embedding of the model within the scenario of Asymptotic Safety, while all main equations are explicitly presented. 

}

\begin{document}

\hoffset = -1cm
\textwidth = 19cm

\maketitle
\flushbottom
\section{Introduction}
The discovery of the Higgs boson at the Large Hadron Collider at CERN \cite{Aad:2015zhl} marked a new era for particle physics, fitting the last missing piece of the Standard Model (SM). The Higgs particle fits into the theoretical framework of Electroweak (EW) interactions, the theory describing the unification of the electromagnetic and weak forces, and is the first fundamental scalar particle ever observed in Nature. It is the latter fact which makes its discovery of particular significance for cosmology too. In fact, scalar particles have been often hypothesised in cosmology to explain observations associated with the physics of the early or the late-time universe, and particularly in the physics of inflation, the speculated rapid expansion of the universe shortly after the Big Bang. 

Higgs inflation \cite{Bezrukov:2007ep, Bezrukov:2010jz} is the theory which assumes that the SM Higgs particle is responsible for the dynamics of the primordial inflationary period. The idea is attractive for more than one reasons. First of all, because it does not invoke any new, hypothetical particle into the theory, but builds up on the known field content of the SM. What is more, it provides with the opportunity of placing constraints on the parameters of the SM at high energies, much higher than the energies current particle accelerators can reach, through the observations of the Cosmic Microwave Background (CMB) radiation. In particular, the parameters of the Higgs potential, such as the quartic Higgs coupling $\lambda$ measured at the EW scale, have to be extrapolated up to inflationary scales using the appropriate Renormalisation Group (RG) equations. 

Together with the Starobinsky model, Higgs inflation is one of  the most successful models according to the recent Planck-satellite data \cite{Ade:2015lrj}. Both models achieve inflation through a modification of the standard curvature sector of General Relativity (GR),
and in fact, they are related through a conformal redefinition of the metric field, with the respective Einstein-frame potentials exhibiting striking similarities \cite{Kehagias:2013mya,Bezrukov:2011gp}. However, this correspondence concerns the classical dynamics of the theories, and the quantum equivalence is more delicate and involved. For examples of an off--shell quantum inequivalence between the two frames we refer to Refs \cite{Kamenshchik:2014waa, Benedetti:2013nya}.

The quantum, scalar and tensor fluctuations of the Higgs coupled to gravity during inflation provide tight constraints on the model's parameters at inflationary scales. In particular, the amplitude of the yet unobserved tensor fluctuations are of the order of the scalar potential, $\sim U(\phi)/\Mp^4$, which for large field values is controlled by the quartic coupling, i.e $U(\phi) \sim \lambda \phi^4$. Extrapolating the SM RG equations up to inflationary scales, the value of $\lambda$ ($\sim 10^{-1}-10^{-2}$) cannot provide the necessary suppression, predicting a tensor spectrum incompatibly large with CMB observations. This problem is circumvented with the addition of a non--minimal coupling between the Higgs and curvature in the action, through a term of the form $\xi \phi^2 R$, with $\xi$ a dimensionless coupling controlling the strength of the interaction. This modification changes the amplitude of the inflationary effective potential to $U(\phi)/\Mp^4 \sim \lambda/\xi^2$, and assuming that $\xi$ is sufficiently large, agreement with observations can be established. In particular, it turns out that in principle $\xi \sim 10^3 - 10^4$, but lower values might be possible in very special cases like the possibility of inflation happening at the critical point \cite{Bezrukov:2014bra,Hamada:2014wna}. 

The non--minimal coupling $\xi$ is the only free coupling in the theory to be fixed by cosmological observations, since the value of the quartic coupling $\lambda$ is predicted by the SM equations, modulo uncertainties in the value of the top-quark mass. At the energy scale where Higgs inflation occurs the effect of quantum-gravitational dynamics cannot be in principle neglected, however during inflation the expectation is that they are sufficiently small. The simple argument behind this assumption is that the large value of $\xi$ is expected to provide a sufficiently high suppression of the quantum corrections due to Higgs and graviton loops during inflation, since the respective propagators receive a suppression by factors of $1/\xi$, remembering that the two fields are kinetically mixed in the Jordan-frame action.

The {\it aim of the current work} is to explicitly study what the Wilsonian functional RG predicts for the quantum corrections of the Higgs non--minimally coupled to gravity at the inflationary regime and beyond, including the effect of gravitons. The formalism employs the Wilsonian idea of calculating quantum corrections, based on an infrared RG scale $k$. As we will see, provided the RG scale is consistently chosen, quantum corrections during inflation can be sufficiently suppressed. Since the framework extends in principle to the non-perturbative regime, we will finally briefly discuss the potential embedding of the model within the Asymptotic Safety (AS) scenario for quantum gravity.

We structure the paper as follows: In Section \ref{sec:Motivation} we very briefly review previous results in the literature and motivate our analysis, while Section \ref{sec:Higgs-intro} lays down the equations governing the classical inflationary dynamics for the theory appropriately adopted to our setup. In section \ref{sec:Quantum-calculation} we calculate the RG flow equation and beta functions describing the renormalisation of Newton's constant, as well as the Higgs' quartic and non--minimal coupling, including quantum gravitational corrections up to leading order using the framework of the Wilsonian functional RG. In Section \ref{sec:QDynamics-higher-energies} we use the previously derived equations to analyse the quantum dynamics during inflation, while Sections \ref{sec:QDynamics-higher-energies} and \ref{sec:Asymptotic-safety} investigate the regime beyond inflation in this context, and the possible connection with the scenario of Asymptotic Safety respectively. Some issues related to the dependence on the choice of gauge and regulator are discussed in Section \ref{sec:Gauge-regulator-etc}. We conclude in Section \ref{sec:Discussion}, while explicit intermediate calculations are kept for the Appendix.  
 
\section{A very brief review of quantum effects during Higgs inflation} \label{sec:Motivation}
The value of the essential couplings of a theory is dictated by experiment at a particular physical scale.
As discussed in the Introduction, for a successful Higgs inflation, the non--minimal coupling has to be set to a quite large value, $\xi \sim 10^3 - 10^4$. 
The important question which arises is how stable the couplings' values are under quantum corrections; in particular, within inflation the latter could in principle spoil the flatness of the effective potential. 

Within an effective field theory approach the term $\xi \phi^2 R$ makes perfect sense as part of a leading-order operator expansion, while operators of mass dimension higher than four are usually related to the violation of tree-level unitarity. The scale at which this is expected to occur has been calculated in Ref. \cite{Bezrukov:2010jz}, where after expanding the action around a flat spacetime, and identifying the potentially dangerous operators, it was found to be $\Lambda \sim \sqrt{\xi} \bar{\phi}$. In Ref. \cite{Bezrukov:2010jz} it is argued that its particular value poses no danger for the model. 

Quantum corrections for the system of a scalar (non-minimally) coupled to gravity have been studied in various settings in \cite{Barvinsky:1993zg, Shapiro:1995yc,Narain-Perc_RG-ST1,Narain:2009gb,Oda:2015sma,Zanusso:2009bs,Shapiro:2015ova}, while the particular case of Higgs inflation has been studied in Refs \cite{Barvinsky:1998rn, Barvinsky:2009ii,Barvinsky:2009fy} employing semi-classical, effective-action methods at 1--loop, as well as in Refs \cite{DeSimone:2008ei, Bezrukov:2009db, George:2013iia, George:2015nza,Salvio:2013rja, Salvio:2015kka} in a standard perturbative context. Ref. \cite{Moss:2014nya} studied the Higgs-inflationary action within the approach of the Vilkovisky effective action, including the effect of gravitons, however the running and dynamics of the couplings was not considered there. \footnote{For a recent alternative scenario for Higgs-type inflation embedded within a quantum gravitational setup see Ref. \cite{Salvio:2014soa}.}

Gravity is well known to be perturbatively non--renormalisable, however, it is a well--working quantum effective theory for energies below the Planck scale. 
Although strong quantum-gravity effects are usually assumed to manifest themselves at the Planck scale, their effect can potentially be important at energies as low as the GUT scale. For Higgs inflation it is expected that for $\xi \gg 1$, the large effective Planck mass will provide an $1/\xi$-suppression to graviton and quantum loops, as it is argued in \cite{Barvinsky:1998rn, Barvinsky:1993zg, Barvinsky:2009ii,Barvinsky:2009fy, DeSimone:2008ei}.

Within the Wilsonian implementation of the functional RG we will employ here, the effect of gravitons will be explicitly accounted for, while the regularisation scheme used, based on an infrared sliding RG scale $k$, is able to capture all types of divergences (power law and logarithmic ones) in the effective action. In this context, for energies below the Planck scale, the usual concepts of effective field theories apply, however, the assumption of Asymptotic Safety allows the extension to the deep UV where the relevance/irrelevance of different operators becomes a prediction of the theory.

\section{The Higgs as the inflaton} \label{sec:Higgs-intro}
In this section we will review the classical dynamics of the model adopted to our context, and introduce the relevant characteristic energy scales involved. 
Considering an excitation $\phi$ of the SM Higgs field around its classical, vacuum expectation value (v.e.v) and rotating to the unitary gauge the corresponding lagrangian can be written as the sum of the following three pieces,
\begin{align}
\mathcal{L}_{\text EW} =  
-\frac{1}{2}(\partial_\mu \phi)(\partial^\mu \phi) - V(\phi) +  \mathcal{L}_{\text{W,Z}} + \mathcal{L}_{\text{ferm.}}+  \mathcal{L}_{\text{Yukawa}}.
\end{align}
The first two terms describe the Higgs sector, while the third one is the gauge part describing the field strengths for the $U(1)_Y$ and $SU(2)$ sectors, associated with the photon ($A_\mu$) and the three vector bosons $W^{\pm}$ and $Z$ respectively. The fermionic part describes the kinetic terms for the fermionic degrees of freedom, while the Yukawa term describes the interactions between the Higgs and the other Standard Model (SM) fields through the usual Yukawa couplings. 
The Higgs potential is defined as 
\be
V(\phi) = v_k+ \frac{1}{2}m_k^2 \phi^2 + \frac{1}{4}\lambda_k \phi^4, \label{Higgs-V}
\ee 
and for $m^2 > 0$ ($m^2 <0$) we are in the symmetric (broken) phase, while $v$ represents the vacuum energy. The index $k$ implies that the corresponding quantity is scale dependent, running under the Renormalisation Group (RG) scale $k$. We will make this notion more precise in Section (\ref{sec:Quantum-calculation}). 

The coupling of the sclar $\phi$ to gravity will be described by the following action
\begin{align}
S & =  \int \sqrt{-g} \; \frac{f(\phi)}{2}R-\frac{1}{2}(\partial_\mu \phi)(\partial^\mu \phi) - V(\phi),  \label{Higgs-action}
\end{align}
with $V(\phi)$ given in (\ref{Higgs-V}), while for Higgs inflation the function $f$ is defined through 
\be
f(\phi) \equiv \Mpk^2 + \xi_k \phi^2. \label{ansatz-f}
\ee
Notice that the Planck mass is allowed to run with energy, and is related to Newton's coupling through $\Mpk^2 \equiv m_{p_k}^2/(8\pi) = 1/(8\pi G_k)$.


For a usual slow-roll inflationary phase to take place, the Higgs potential is required to be sufficiently flat, in which case the field starts from an unstable vacuum phase, and after a period of slow roll, it evolves towards its true minimum. 
It is instructive to transform to the Einstein frame, where the fields' kinetic terms diagonalise, defining the conformal transformation to a new metric $\tilde{g}_{\alpha \beta}$ as 
\be
\tilde{g}_{\alpha \beta} = \frac{f(\phi)}{\Mp^2} g_{\alpha \beta},
\ee
with $M_{p_0}^2 \equiv m_{p_0}^2/(8\pi) = 1/(8\pi G_0)$, the Planck mass as measured at solar-system scales.
The following field redefinition will yield a canonically normalised scalar $\chi$, 
\be
\left(\frac{d \chi}{d \phi} \right)^2 = \frac{\Mp^2}{f(\phi)} + \frac{3}{2} \Mp^2 \left( \frac{f'(\phi)}{f(\phi)}\right)^2 
= \frac{x}{1 + \xi_k  x     \hat{\phi}^2} +    \frac{6 \xi_k^2 x^2  \hat \phi^2}{ \left( 1 + \xi_k  x  \hat {\phi} ^2 \right)^2}, \label{chi-phi-redefinition}
\ee
where in the last step we used (\ref{ansatz-f}) to substitute for $f$, and defined the following useful quantities
\begin{align}
\hat \phi \equiv \phi / \Mp, \; \; \; x \equiv x(k) \equiv \frac{\Mp^2}{\Mpk^2} \equiv \frac{G_k}{G_0}. \label{def:hatphi-x}
\end{align}
The quantity $x \equiv x(k)$ modifies most of the standard inflationary relations and has to be evaluated for the value of the coupling $G(k)$ during inflation, i.e $G(k = k_{\text{inflation}})$. For $x(k) = 1$, one recovers the standard results. It turns out from the analysis of Section \ref{sec:Q-dynamics-inflation} that during the inflationary regime it will be $x(k) \simeq 1$ ($G_{k} \simeq G_0 = \text{constant}$) to very good accuracy.
The Einstein-frame action reads as
\be
\tilde S =  \int \sqrt{-\tilde{g}} \; \frac{\Mp^2}{2}  \widetilde{R}-\frac{1}{2}(\partial_\mu \chi)(\partial^\mu \chi) - U(\chi),
\ee
with $U$ defined as
\be
U[\phi(\chi)] \equiv \Mp^4 \cdot \frac{V[\phi(\chi)]}{(\Mpk^2 + \xi_k \phi(\chi)^2)^{2}}.
\ee
The potential $U$ depends implicitly on the Einstein-frame scalar $\chi$, a choice which is convenient for the calculation of inflationary observables. Inflation will occur at sufficiently high energies, where  $8 \pi G_k \xi_k \phi^2 \equiv  x \xi_k  \hat \phi^2 \gg 1$, and $V(\phi) \simeq (\lambda_k/4) \phi^4$. In this regime, (\ref{chi-phi-redefinition}) can be  integrated to give
$
\chi(\phi) \simeq \sqrt{\frac{3}{2}} \Mp \cdot \log(1+ \xi_k \hat \phi^2 x),
$
leading to the following explicit form of $U(\chi)$
\be
U(\chi) = \Mp^4  \cdot  \frac{\lambda_k}{ 4 \xi_k^2}  \cdot  \left( 1 - e^{- \sqrt{2/3}\cdot   \frac{\chi}{\Mp}}\right)^2. \label{potential:EF}
\ee
For $\chi/\Mp \gg 1$ the potential approaches a constant value $U(\chi) \simeq \Mp^2   \cdot   \frac{\lambda_k}{4 \xi_k^2}$ corresponding to the slow--roll regime.

Varying the action with respect to the metric, and evaluating on a flat, Friedmann-Lemaitre-Robertson-Walker (FLRW) spacetime, in the slow-roll regime the Friedman equation becomes
\be
H^2 \simeq x(k)\cdot \frac{U(\phi)}{ 3 M_{p_0}^2},
\ee
with the Hubble parameter $H$ defined as $H(t) \equiv \dot a(t)/a(t)$, $a(t)$ being the scale factor, $t$ the cosmic time in the Einstein frame and $x(k)$ defined in (\ref{def:hatphi-x}). 
The slow-parameter $\epsilon$ is defined in the standard way as
\begin{align}
\epsilon &\equiv - \frac{\dot H}{H^2} \simeq  \frac{\Mp^2}{2} \left( \frac{U_{, \chi}}{U} \right)^2 =   \frac{\Mp^2}{2} \left(\frac{V_{, \phi}}{V} \frac{1}{\chi_{,\phi}} \right)^2 = \frac{4}{3}  \cdot  \frac{1}{x^2}  \cdot   \frac{1}{\xi^2 \hat \phi^4}, \label{epsilon}
\end{align}
with $, \equiv \partial / \partial \phi$, while the number of e-foldings $N$ between $\phi_i \to \phi_f$ is given by 
\begin{align}
N = \int_{\phi_f}^{\phi_i}\frac{1}{M_{p0}^2} \frac{V}{V_{, \phi}} \left( \chi_{,\phi} \right)^2 
    = \frac{3}{4}  \xi\cdot  x \cdot \left( \hat \phi_i^2 -  \hat \phi_f^2 \right) +  \frac{3}{4}    \log \left( \frac{1 + \xi \cdot x \cdot \hat \phi_{f}^2}{1 +\xi \cdot x \cdot \hat \phi_{i}^2} \right). \label{N}
\end{align}
For slow--roll inflation it is $\epsilon \ll 1$, which implies that inflation starts for field values around $\phi \gtrsim \Mp/{\sqrt{\xi}}$, where for simplicity we set $x(k) = 1$. 
To find the starting value of the field $\phi_i$, we evaluate expression (\ref{N}) at the required number of e-foldings $N =N_0$, before the end of inflation, while the condition $\epsilon \simeq 1$ in (\ref{epsilon}) will yield the value of $\phi = \phi_f$ at the end of inflation respectively. We find that,
\begin{align}
& \frac{\phi_i}{\Mp} \simeq \frac{1}{\sqrt{x}}   \cdot   \frac{1}{\sqrt{\xi}}  \cdot   \left( \frac{4}{3}N_0 + \frac{2}{\sqrt{3}} \right)^{1/2}, \label{phi-i} \\
& \frac{\phi_f}{\Mp} \simeq \left( \frac{4}{3} \right)^{1/4} \cdot  \frac{1}{\sqrt{x}} \cdot \frac{1}{\sqrt{\xi}}. \label{phi-e}
\end{align}
For $x(k) = 1$, $N_0 = 55$, one finds $\phi_i \simeq 8.631 \Mp/\sqrt{\xi}$ and $\phi_f \simeq  1.075 \Mp/\sqrt{\xi}$ respectively.
%
\footnote{Notice that these are the field values in the Jordan frame. The corresponding ones in the Einstein frame have to be translated through $\chi =\chi(\phi)$ given a little before (\ref{potential:EF}).}
%
%

The vacuum fluctuations of the inflaton produce a spectrum of scalar and tensor perturbations, which' amplitudes evaluated at horizon crossing at the pivot scale $k_{ \text{pivot}} = 0.002 \, {\text{Mpc}}^{-1}$ read as \cite {Lyth:2009zz}
\begin{align}
P_{\text{S}} = \frac{1}{24 \pi^2} \frac{1}{\Mp^4} \frac{U[\phi(\chi)]}{\epsilon}
\simeq \frac{1}{128 \pi^2}  \cdot  x^2  \cdot  \lambda  \cdot  \hat \phi^4,
\; \; \; \; \; \; P_{\text{T}} = \frac{128}{3} \frac{U[\phi(\chi)]}{\Mp^4} \label{Spectra},
\end{align}
with the field value in the last relation assumed to be $\phi = \phi_i$. 
With the aid of (\ref{phi-i}), and assuming the observed value for the amplitude of the scalar fluctuations evaluated at horison crossing, $P_{\text{S}} = P_{\text{S(obs.)}}$ as required from CMB observations, we can work out the relation between the couplings $\lambda$ and $\xi$ during inflation as \footnote{Unless otherwise stated, we will be assuming $N_0 = 55$.}
\be
\xi_{\text{inflation}} \simeq  \frac{1}{\sqrt{128} \pi}  \cdot  \frac{1}{\sqrt{P_{\text{S(obs.)}}}}  \cdot \left( 4N_0 +2\sqrt{3} \right)  \cdot  \sqrt{\lambda}. \label{xi-lambda-relation}
\ee
The coupling $\xi$ in (\ref{xi-lambda-relation}) depends on the cosmological parameters such as the number of e-foldings and the amplitude of scalar fluctuations, but also on SM parameters such as the top quark/Higgs mass at the EW scale which enter implicitly through the coupling $\lambda$, so that we can write 
\be
\xi_{\text{inflation}} \equiv \xi[N_0, P_{\text{S(obs.)}}; \; \lambda_{(\text{EW})}, M_{t(\text{EW})}, \cdots],
\ee
with the index $(\text{EW})$ standing for the value at the EW scale. A typical value for the coupling $\lambda$ at inflationary scales is $\sim 10^{-2}$ yielding $\xi \sim 10^{3}$ (see Appendix \ref{sec:EWTOINFL} for a realistic evaluation).

In overall, the classical dynamics of the model define two characteristic scales, the typical value of the (Jordan-frame) scalar field at the end of inflation, $\phi_{f} \sim \Mp/\sqrt{\xi}$, and the Hubble scale during inflation, $H \sim \Mp/\xi$. These in turn define three characteristic energy regimes. In the particular setup of this work, there is yet one more scale, the sliding RG cut--off $k$, representing the typical energy (coarse-graining) scale of the system. Its value and connection with the standard scales during inflation will be discussed in Section \ref{sec:Q-dynamics-inflation}.


\section{Quantum gravitational corrections and running during inflation} \label{sec:Quantum-calculation}

\subsection{The setup} \label{sec:Quantum-calculation-Setup}
Our final  aim is to understand what the Wilsonian functional RG predicts for the quantum (gravitational) corrections for the model, under certain assumption which we describe below. We will first introduce the basic concepts and tools needed for the subsequent analysis, and also remind that explicit calculations and formulae are presented in the Appendix. 

Let us start with some theory and its bare action $S[\varphi^A]$ which depends on a set of fields $\{ \varphi^A \}$, with $A, B$ generalised field/spacetime indices.
Formally, the construction of the associated effective action starts with the generating functional of the connected Green's functions, 
\be
W[J] \equiv \log Z[J] = \log \int \mathcal{D}\varphi_A \exp \left[S[\varphi_A] + \int J^B(x) \cdot \varphi_B(x) \right],
\ee
with $J^{B}(x)$ standing for the sources. From the functional $W[J]$ one can define the expectation value of the fields as $\Phi^B(x) \equiv \braket{\varphi^B} = \delta W[J^A]/\delta J^B$. The effective action $\Gamma$ \footnote{For a rigorous discussion on reconstructing the microscopic bare action from the Wilsonian effective action within the functional RG and Asymptotic Safety see \cite{Manrique:2008zw, Morris:2015oca}.} is then introduced through a change of field variables by means of a Legendre transform, where the sources in $W[J]$ are traded for the fields $\Phi^A$ as
\be
\Gamma[\Phi] = \int_x \Phi^B \cdot J_{B}[\Phi] - W\left[ J^A[\Phi] \right].
\ee
It is well known that the 1-loop corrections of the theory are intimately connected to the (Eucledian) effective action $\Gamma$ through the following relation 
\be
\Gamma^{\text{1-loop}} = \frac{1}{2} \text{Tr} \log S^{(2)}. \label{Gamma-1-loop}
\ee
The quantity $S^{(2)}$ stands for the inverse bare propagator defined as $S^{(2)} \equiv \nicefrac{\delta^{2} S}{\delta \Phi^A \delta \Phi^B}$, and possible index structure is understood, while  ''Tr" stands for summation over spacetime, internal indices and momenta. 
The trace over momenta of the kinetic operators leads to an in principle divergent result which requires some sort of regularisation. There are different types of regularisation schemes, each with its own advantages and disadvantages, two of the most popular being dimensional regularisation and a physical cut--off respectively. The Wilsonian approach suggests a continuous integrating out of momenta,  shell-by-shell in momentum space. The functional RG we will employ here, implements this idea by invoking an IR regulator, denoted as $R_k$, in turn built out of an infrared, dimensionfull cut-off $k$. Its generic form is constrained by certain conditions, see \cite{Gies:2006wv, Reuter:2012id, Pawlowski:2005xe, Litim:2008tt} for a discussion. Above ideas lead to the concept of the Wilsonian, or average effective action $\Gamma_k[\Phi^A] $ defined as
\be
\Gamma_k[\Phi^A] = \Gamma[\Phi^A]  - \Delta S_k[\Phi^A], \label{Wilson-Eff-Action}
\ee
with $\Delta S_k[\Phi^A] \equiv \frac{1}{2}\int \Phi^{A} \cdot R_{k AB} \cdot \Phi^{B}$. By construction, the regulator $R_k$ employs an infrared regularisation, suppressing fluctuations with momenta $p^2 < k^2$, while integrating out those with $p^2 > k^2$. We will get back to the particular choice of the regulator $R_k$ later. In view of (\ref{Wilson-Eff-Action}), the suppression term $ \Delta S_k$ amounts to the modification of the full inverse propagators $\Gamma^{(2)}$ according to
$
\Gamma^{(2)} \to \Gamma^{(2)} + R_k,
$
and it is understood that $R_k$ should carry the same tensor structure with $\Gamma^{(2)}$. The cut-off scale $k$ is interpreted as the typical energy scale, or equivalently, $1/k$ defines the typical physical lengthscale of the system one is interested in. 

It can be then shown that the average effective action (\ref{Wilson-Eff-Action}) satisfies an Exact Renormalisation Group Equation (ERGE) \cite{Wett_RG-Eq, Morris:1994ie}
\be
\partial_{t} \Gamma_k =  \frac{1}{2} \text{Tr}  \left( \Gamma^{(2)} + R_{k} \right)^{-1}\cdot \partial_{t} R_{k}, \label{ERGE}
\ee
with $\partial_t \equiv k \partial_k \equiv k \partial/\partial k$.
This last equation will play an important role for our quantum analysis. 
For $\Gamma^{(2)} \to S^{(2)}$ equation (\ref{ERGE}) connects with the standard 1--loop result (\ref{Gamma-1-loop}), its applicability though extends beyond the perturbative regime. Exact solutions within a gravitational context are almost impossible, and some sort of approximation has to be invoked. Notice also that, equation (\ref{ERGE}) is an in principle off--shell equation, which makes any results derived from it dependent on the gauge, while the use of approximations like truncated actions leads to a dependence on the regularisation scheme. In the context of scalar--tensor theories another subtlety arises regarding the choice of the conformal frame, with off--shell corrections not in principle expected to match in different frames, as explicitly shown in Refs \cite{Kamenshchik:2014waa, Benedetti:2013nya}. 

Let us summarise the {\it basic assumptions for the quantum analysis} as follows:\\
\vspace{0.08cm}
1. \; We will assume that the effective action takes the form suggested by (\ref{Higgs-action}), ignoring higher-order operators, while the calculation will be performed in the Jordan frame. Notice that below, we might sometimes drop the index ''$k$" for the running couplings for simplicity. 
\vspace{0.08cm}
\\
2. \;  The usual background field method for the decomposition of the fields into a background and fluctuating part in a Euclidean signature will be employed. For the background-value of $\phi$ we will assume that $\partial_\mu \bar{\phi} = 0$, while the background spacetime will be a Euclidean de Sitter. The trace over momenta in (\ref{ERGE}) will be performed with the use of a heat kernel expansion. 
\vspace{0.08cm}
\\
3. \; We will consider the quantisation of the gravity-scalar sector only, hence Yukawa and gauge interactions will not be accounted for in the calculation. 
We will also assume that the quartic coupling $\lambda$ retains a positive value, since the possible instability of the Higgs potential poses an important problem which deserves its own study. We briefly discuss this issue in Appendix \ref{sec:EWTOINFL} numerically solving the 1-loop SM RG equations. 
\\
4. \; We will perform the calculation in the popular choice of the de Donder gauge which significantly simplifies the technical analysis. We comment on the gauge and regulator choice in Section \ref{sec:Gauge-regulator-etc}.
\vspace{0.08cm}
\\

\subsection{The calculation} \label{sec:Quantum-calculation}
We can now start with the calculation within the ERGE. Our goal is to evaluate (\ref{ERGE}) for the action ansatz of (\ref{Higgs-action}) and under the assumptions described earlier. \footnote{Notice that similar calculations within scalar-tensor theories have been performed in \cite{Narain-Perc_RG-ST1,Narain:2009gb,Oda:2015sma} using different field decomposition and evaluating the flow equations in Landau gauge and optimised cut--off respectively, as well as in \cite{Zanusso:2009bs} around a flat background including fermions, and more recently, in \cite{Shapiro:2015ova} within a semi-classical setting.}
The gravity sector has the usual diffeomorphism gauge symmetry, which we will fix through the introduction of a gauge-fixing term. The Wick-rotated and gauge-fixed effective action ansatz then reads as
\be
\Gamma[g_{\mu \nu}, \phi, C^{\mu}, \bar{C}^{\nu} ] =  -\int \sqrt{-\bar g} \; \left( f_k(\phi)R-\frac{1}{2}g^{\mu \nu}(\partial_\mu \phi)(\partial_\nu \phi) - V_k(\phi) \right) + S_{ghost} + S_{GF}. \label{Gamma-Higgs}
\ee
The terms $S_{GF}$ and $S_{ghost}$ stand for the gauge-fixing and ghost sector respectively, while $ C^{\mu}, \bar{C}^{\nu}$ denote appropriate ghost and anti-ghost fields. We define them explicitly below. The indices $k$ remind us that the quantities stand for the renormalised ones, running under the RG scale $k$.  
The gauge-fixing term is defined as follows
\begin{align}
& S_{GF} = \frac{1}{2\alpha} \int \sqrt{g} \, f(\phi) \bar g^{\mu \nu} \chi _{\mu} \chi_{\nu}, \; \; \; \text{with} \; \; \;  \chi^{\mu} = \nabla_{\nu} h^{\mu \nu} - \frac{\beta + 1}{4} \bar g^{\mu \nu} \nabla_{\nu} h,
\end{align}
which depends on the two real parameters $\alpha$ and $\beta$. Two of the most popular choices in the literature correspond to $\alpha = \beta = 1$ (de Donder gauge), and $\alpha \to 0$ (Landau-type gauge) respectively. For our analysis, we will choose the first with $\alpha=\beta = 1$, which simplifies the calculation. 
Now, the introduction of the gauge-fixing term requires the introduction of appropriate ghost and anti-ghost fields which can be calculated by replacing the gauge vectors $u^\mu$ in the gauge transformation of the combined metric $\mathcal{L}_u(g_{\mu \nu}) =\mathcal{L}_u(\bar g_{\mu \nu} + h_{\mu \nu}) = u^{\rho}\partial_{\rho} g_{\mu \nu} + \partial_{(\mu}u^{\rho}g_{\nu) \rho}$, by the ghost $C^\mu$.
Then, following the standard Fadeev--Poppov procedure the ghost term can then be shown to take the form
\be
S_{ghost} = -\int d^{4}x  \sqrt{g} \bar{C}_{\mu} \left( \delta^{\mu}{}_{\nu}(- \Box)  -  R^{\mu}{}_{\nu} \right) C^{\nu},
\ee
with $C^\mu, \bar C^\mu$ denoting the ghost and anti-ghost fields respectively, and $\Box \equiv \bar{g}^{\mu \nu} \bar{\nabla}_\mu \bar{\nabla}_\nu$.

%
Expanding the effective action (\ref{Gamma-Higgs}) up to second order in the field's fluctuations under (\ref{Field-expansion}) we calculate the Hessians $\Gamma_{k}^{(2)}$, which' inversion yields the different propagator entries appearing in (\ref{ERGE}) (or (\ref{Gamma-1-loop})). The explicit expressions are given in Appendix \ref{Appendix:Expl-rel}. To this end, we employ the background field method by considering the following split between a background piece (denoted by an overbar) and a fluctuating part as \footnote{We should notice here that different parametrisations have been employed in the literature, such as the exponential one, see e.g Ref. \cite{Nink:2014yya}.}
\be
g_{\mu \nu} =\bar g_{\mu \nu} + h_{\mu \nu}, \; \; \;  h_{\mu \nu} =  \hat{h}_{\mu \nu} + \frac{1}{4} \bar g_{\mu \nu} h, \; \; \; \; \; \; \; \phi = \bar \phi + \delta \phi, \label{Field-expansion}
\ee
with $\bar g_{\mu \nu}$ describing the background spacetime metric and the trace--free (denoted with a hat) and trace components of the metric fluctuation satisfying $\bar g^{\mu \nu} \hat{h}_{\mu}{}_{\nu} = 0$, $h \equiv  \bar g^{\mu \nu} h_{\mu \nu}$. Derivatives will be constructed with the background metric field, and we shall drop the overbar from them for notational convenience. We notice that the fluctuating fields $h_{\mu \nu}$ and $ \delta \phi$ are assumed to be the corresponding average fields, i.e $h_{\mu \nu}(x) \equiv \braket{h_{\mu \nu}(x)}, \delta \phi(x) \equiv \braket{\delta \phi(x)}$. Ideally, one would like to keep the background field variables unspecified, however this can be technically unpractical and lead to results of very high complexity; we refer to \cite{Bridle:2013sra, Becker:2014qya,Demmel:2014sga} for recent discussions within a functional RG context. In this work, we will assume the family of constant backgrounds of a four-dimensional Eucledian de Sitter ($S_4$) with 
\be
\bar R, \bar \phi =  \text{constant}.
\ee

In the quadratic part of the expanded action, the different interaction vertices appearing are the effective graviton and Higgs self interactions, as well as the momentum-dependent cross-vertex between the scalar and the metric, due to the non-minimal coupling (see Appendix \ref{Appendix:Expl-rel}).
On $S_4$, the kinetic part of it consists of a minimal operator which is regularised with the introduction of the one-parameter regulator $R_k \equiv R_k(-\Box; r)$, through the modification \cite{Reuter:1996cp}
\be
\Gamma_{k}^{(2)}(-\Box) \to \Gamma_{k}^{(2)}(-\Box)  + R_k(-\Box;r).
\ee 
This way, the eigenvalues of $-\Box$ less than $k^2$ are suppressed, while integrated out otherwise. As the cut--off is continuously moved, the integrating out of modes is performed shell-by-shell in momenta \cite{Gies:2006wv, Reuter:2012id, Pawlowski:2005xe, Litim:2008tt}. As regards the particular choice of regulator function, we choose an 1--parameter version of the optimised regulator \cite{Litim_Opt-CO1} 
$
R_{k}\left( -\Box \right) \equiv \left( r\cdot k^2 - (-\Box) \right) \cdot \Theta \left(r\cdot k^2 - (-\Box) \right),  
$
which will allow for an explicit computation of the the momentum integrals.
The real and positive parameter $r$ defines a family of regulator functions, with the standard, ''optimal" case corresponding to $r=1$. It will serve as a book-keeping parameter which we will use as a test of the regulator-dependence of the main results. 

The sum over the eigenvalues of the operators appearing in the 1--loop-type trace on the right-hand side of the ERGE is traced by 
means of an asymptotic heat kernel expansion. Assume an operator $\Delta = -\Box \delta^{A}{}_{B} + U^{A}{}_{B}$, with $\Box \equiv \bar{g}^{\mu \nu} \bar{\nabla}_{\mu} \bar{\nabla}_{\nu}$, $\delta^{A}{}_{B}$ the identity matrix in field space, and $U^{A}{}_{B}$ a potential-type term depending on the background value of the fields and their derivatives. Then, in four dimensions the heat-kernel expansion of $\Delta$ reads
\be
\text{Tr} e^{-s \Delta } = \left( \frac{1}{4 \pi s} \right)^{2} \int d^{4}x \sqrt{\bar{g}} \left( \text{tr} a_0 + \text{tr}a_2 s +  \text{tr}a_4 s^2 + \ldots  \right), \label{HK-exp-Delta}
\ee
where the parameter $s$ is assumed to be sufficiently small, and $\text{tr}$ sums over internal indices. The coefficients $a_i$ depend on the background geometry, with each term in the expansion capturing different types of divergences, in particular quartic ($a_0$), quadratic ($a_2$) and logarithmic ($a_4$) divergences respectively \cite{Vassilevich:2003xt, Reuter:1996cp, Reuter:2007rv, Codello:2008vh}. Formally the expansion (\ref{HK-exp-Delta}) is valid as long as $\bar{R}/k^2 < 1$, i.e capturing fluctuations with wavelengths smaller than the radius of curvature.

Evaluating the trace in the ERGE (\ref{ERGE}), the flow equation for $\Gamma_k$ turns out to organise in the following form
\begin{align}
k^4 V \cdot \partial_t \Gamma_k = \mathcal{F}_0 + \mathcal{F}_{1} \cdot  \frac{\partial_t f}{f} + \mathcal{F}_{2}\cdot \frac{\partial_t f'}{f'}, \label{Flow:Effective-action-main}
\end{align}
with primes here denoting derivatives with respect to $\tilde{\phi} \equiv \phi/k$, $\partial_t \equiv k \partial_k$ and $V$ is the volume of $S_4$. The dimensionless quantities $\mathcal{F}$ depend non--trivially on the fields, couplings, and regulator parameter  $\mathcal{F} = \mathcal{F}[R, \phi ; f, V; r]$, with their form explicitly given in (\ref{F-explicit}) of the Appendix. As can be seen from their explicit expressions, the functionals $\mathcal{F}$ depend up to second order derivatives of $f$ and $V$ with respect to $\phi$, as expected. The flow described by (\ref{Flow:Effective-action-main}) is particularly involved, however, its 1--loop expression simplifies considerably, and is also explicitly presented in Appendix \ref{Appendix:Flow-1loop}. From (\ref{Flow:Effective-action-main}), expanding around $\bar{R}/k^2 = 0$, $\bar{\phi}/k = \tilde{\phi}_*$, and projecting out on the different operators in the effective action one gets the flow of the two scalar potentials as,
\begin{align}
& \partial_t f(\phi) =  \mathcal{F}_f[\tilde{\phi}_*; g_j, \partial_t g_j; r] , \; \; \; \; \partial_t V(\phi) =  \mathcal{F}_V[ \tilde{\phi}_*; g_j, \partial_t g_j;  r],  \label{Flow:general}
\end {align}
with $\mathcal{F}_f$ and $\mathcal{F}_V$ corresponding to the projection of $\mathcal{F}$ on the curvature and scalar potential operators respectively. 
In turn, projecting out on the individual operators in $f$ and $V$ one can extract the running of the individual coupling constants. Notice that the evaluation of the ERGE generates higher-order terms in curvature/scalar field which we neglect in view of our original action ansatz. 

\subsection{The structure of the beta functions}
During slow-roll inflation, the scalar field acquires a large vacuum energy, $\phi = \phi_{*} \gtrsim \Mp/\sqrt{\xi}$, and we therefore consider an expansion of $V$ around this v.e.v as
\be
V_k(\phi) = v_k +  \lambda_k (\phi^2 - {\phi_{*}}_k^2)^2, \label{potential-broken}
\ee
with $v_k$ representing a cosmological constant-type term. The function $f$ will have the form of (\ref{Higgs-action}).
At this stage it is convenient to introduce dimensionless fields and couplings, measured in units of the cut--off $k$, 
 \be
\tilde{\phi}_{*k} \equiv \phi_{*}(k)/k, \; \; \;  \tilde{g}_i \equiv g_i(k)/k^n, 
\ee
where $n$ is the coupling's canonical dimension. Under the ansatz (\ref{potential-broken}), from the flow equation (\ref{Flow:general}) one can extract an autonomous system of (non--perturbative) beta functions
\be
k \partial_k \tilde{g}_i  = \beta_{\tilde{g}_i}(\tilde{g}_j) \equiv (- n+ \eta_{g_{i}}(g_j))\cdot \tilde{g}_i ,
\ee
with $\eta$ the anomalous dimension of the coupling.
Due to the appearance of RG-derivatives on both sides of the equation (\ref{Flow:Effective-action}), the resulting expressions are very involved, but they simplify significantly in the 1--loop approximation where the RG-derivatives on the r.h.s of (\ref{Flow:Effective-action}) are switched off. The explicit expressions in the 1--loop approximation are presented in the (\ref{Appendix:beta-G-1L})--(\ref{beta-v-Appendix}) of the Appendix, while the beta functionals for $\widetilde G$ and $\xi$ are also explicitly given in the limit $\tilde{\phi}_{*}, \tilde{v} \to 0$ without any further approximation assumed.

The general structure of the equations in the 1--loop approximation reads \footnote{To form a simple basis for our discussion we will choose the ''optimised" cut--off parameter $r=1$. We briefly comment on the gauge and regulator dependence in Section \ref{sec:Gauge-regulator-etc}. We also neglect the contribution of $v_k$ in the beta functions other than its own one.}
\be
\beta_{i} \equiv \beta_{i}^{(0)}  + \beta_{i}^{(\text{grav.})}  = \beta_{i}^{(0)}  +  \Omega^{-m_i}  \cdot \sum_{n \geq 1} \mathcal{B}^{(n)}(\tilde{\phi }_*, \xi) \cdot (\tilde{G}\tilde{\phi }_{*}^{2})^n, \label{Beta-general-struct}
\ee
where the coefficients $\mathcal{B}^{(n)}$ and exponents $m_i$ can be read off from (\ref{Appendix:beta-G-1L})--(\ref{beta-v-Appendix}), together with the definitions
\begin{align}
\Omega \equiv \left(16 \pi  \mu  \xi  \tilde{G} \tilde{\phi }_*^2  + 8 \pi (9 \xi +1) \xi \tilde{G} \tilde{\phi }_*^2   + 2\mu +1\right), \; \; \; \mu \equiv \lambda \tilde{\phi }_*^2. \label{Def:OMega-mu}
\end{align}
The way we split the contributions in the beta functions (\ref{Beta-general-struct}) is such that the terms $\beta_{i}^{(0)}$ reduce to the standard perturbative results in the limit $\tilde{\phi }_* \to 0$, while $\beta_{i}^{(\text{grav.})}$ conventionally denote the gravitational corrections to them. This is only conventional, since during inflation the v.e.v $\tilde{\phi }_*$ is in fact related to the non-minimal coupling to gravity $\xi$.
The quantity $\Omega$ appears as a result of the kinetic mixing between the graviton and scalar in the action, and becomes $\Omega = 1$ for $\tilde{\phi }_*  \to 0$, but for sufficiently large $\tilde{\phi }_*$ and $\xi$ it provides a sufficiently high suppression to the different terms in (\ref{Beta-general-struct}). The origin of the non-standard terms $\sim \tilde{G}\tilde{\phi }_* \xi$ is also similar; these terms appear after expanding the non--trivial propagator entries in the ERGE around the v.e.v of the scalar under the particular ansatz for $f$ and $V$ ((\ref{Higgs-action}) and (\ref{potential-broken}) respectively), and obviously, they vanish for $\tilde{\phi }_*  \to 0$. These terms are an immediate result of the scalar's non--zero v.e.v., introducing appropriate threshold effects; it is
\be
\tilde{\phi}_* \equiv \phi_*/k \gg 1,
\ee
for v.e.v values much larger than the cut--off scale $k$, and opposite otherwise. The first case is expected to occur during inflation. The actual estimate of the value of $\tilde{\phi}_* \equiv {\phi}_*/k$ depends on the estimate of the cut--off $k$ for the energy regime of interest. This will be discussed explicitly in Sections \ref{sec:Q-dynamics-inflation} and \ref{sec:QDynamics-higher-energies}. 
In general, for $\widetilde{G}, \tilde{\phi }_* \to 0$ one recovers the {\it standard, perturbative expressions} for the beta functions. 

The beta functions for a non-zero v.e.v $\tilde{\phi }_* \neq 0$, according to (\ref{Beta-general-struct}) (see also (\ref{Appendix:beta-G-1L})--(\ref{beta-v-Appendix}) of the Appendix), read as 


\begin{align}
\beta_{\widetilde{G}} = 2 \widetilde{G} +   \frac{1}{24\pi} \cdot \frac{14 \xi  + 240 \mu^2  +  230 \mu  -  55}{( 1 + 8 \pi \xi \widetilde{G} \tilde{\phi}_{*}^2 )^2 \cdot \Omega^2} \cdot \widetilde{G}^2 + \mathcal{O}\left(\frac{\widetilde{G}^3}{(1 + 8 \pi \xi \widetilde{G} \tilde{\phi}_{*}^2)^2  \Omega^2} \right)\label{beta-G-Landau-vev},
\end{align}
%
%
\begin{align}
\beta_{\xi} = 
\frac{1}{64 \pi^2} \cdot \frac{\lambda \left( 28 \xi + 10 \mu + 5 \right)}{ \Omega^3} +  \beta_{\xi}^{(\text{grav})}  , \label{beta-xi-Landau-vev}
\end{align}
%
%
\begin{align}
 \beta_{\lambda} =
\frac{21}{16 \pi ^2 } \cdot \frac{ \lambda ^2}{( 1 + 8 \pi \xi \widetilde{G} \tilde{\phi}_{*}^2 ) \cdot \Omega^3} +  \beta_{\lambda}^{(\text{grav.)}} \label{beta-lambda-Landau-vev},
%
%
\end{align}
with $\Omega$ and $\mu$ defined in (\ref{Def:OMega-mu}).
From $ \beta_{\xi}$ and $ \beta_{\lambda}$ we can also derive an expression for the fractional running of the amplitude of scalar fluctuations $\sim \lambda/\xi^2$. Keeping only the $\beta_{i}^{(0)}$ terms, we find
\begin{align}
\partial_t \left(\frac{\lambda}{\xi^2} \right)  & \simeq  \frac{\lambda}{\xi^2}  \cdot  \left( \frac{\beta_{\lambda}^{(0)}}{\lambda} - \frac{2\beta_{\xi}^{(0)}}{\xi} \right) 
 \simeq \frac{1}{16 \pi^2} \cdot \frac{\lambda^2}{ \xi^2 \Omega^3 } \left( \frac{21}{1+8 \pi \widetilde{G} \tilde{\phi}_*^2} - \frac{28 \xi + 10 \mu }{2 \xi}  \right) 
 \label{beta-omega-Landau-vev}.
\end{align}
The RG equations for $\phi_*$ and $v$ in (\ref{potential-broken}) can be found in (\ref{beta-phi-Appendix}) and (\ref{beta-v-Appendix}) of the Appendix.

For $\tilde{\phi}_*, \widetilde{G} \to 0$, the terms $\propto  \lambda, \xi\cdot \lambda$ on the r.h.s of the beta function for $\xi$, equation (\ref{beta-xi-Landau-vev}), are in qualitative agreement with those found in the context of the conformal anomaly \cite{Buchbinder:1992rb}, and they tend to increase $\xi$ with the cut--off scale, with $\xi$ admitting the usual logarithmic running. In a similar way, the beta function for $\lambda$, equation (\ref{beta-lambda-Landau-vev}), consists of the standard $\lambda^2$-term leading to logarithmic running and an irrelevant Landau pole at very high energies.

Let us briefly comment on the renormalisation of Newton's coupling. For the purpose of this discussion we re-write (\ref{beta-G-Landau-vev}) as
\be
\partial_t \widetilde{G}  = \left( 2 + \eta_G\right)\widetilde{G},
\label{beta-G-Landau2}
\ee
with $\eta_{G} \equiv  -  Z_G^{-1}\partial_t Z_G$, $Z_G^{-1} \equiv 16 \pi G(k)$.
{A negative anomalous dimension $\eta_G$ will tend to reduce $\widetilde{G}$ and eventually lead it to a UV fixed point} as $k \to \infty$, where $\eta_G = -2$. This lies in the heart of Asymptotic Safety which we discuss in Section \ref{sec:Asymptotic-safety}. On the other hand, $\eta_G > 0$ will have the opposite effect leading the coupling to increasingly large values with increasing $k$. This is an unwanted behaviour if the theory is to posses a well-behaved high-energy regime. 
%


\section{Quantum dynamics during inflation} \label{sec:Q-dynamics-inflation}
In the RG equations (\ref{beta-G-Landau-vev})--(\ref{beta-omega-Landau-vev}) the threshold effects due to the v.e.v of $\phi$ appear through $\tilde{\phi }_* \equiv \phi _*/k$. 
Depending on its value, we distinguish the large- and small- field regime where $ \tilde{\phi }_* \gg 1$ and $\tilde{\phi }_* \ll 1$ respectively. The first case is expected to  occur during inflation, remembering that the scalar acquires a large v.e.v with 
\be
\phi_* \gtrsim \Mp/\sqrt{\xi}.
\ee 

To estimate the value of $\tilde{\phi }_*$, one needs an estimate of the infrared cut--off $k$ during inflation. An important point to make is that the prescription for the interpretation of the infrared cut--off $k$ in this context depends on the particular physical setup. In general, $k$ represents the coarse-graining scale, or the typical energy scale of the physical system (see \cite{Wein, Reuter:2005kb, Guberina:2002wt,Shapiro:2004ch,Babic:2004ev,Domazet:2010bk,Bonanno:2001xi,Bonanno:2001hi,Bonanno:2009nj,Bonanno:2010bt,Bonanno:2012jy,Contillo:2011ag,Frolov:2011ys, Cai:2011kd, Hind-Litim_Rahme, Hindmarsh:2012rc,Koch:2010nn,Koch:2014joa,Gonzalez:2015upa,Bonanno:2015fga}). Quantum fluctuations during inflation are of the order of the cosmic horizon $H^{-1}$, suggesting the coarse-graining scale to be of the same order, i.e $k \sim H$. This is the choice employed in \cite{Wein, Reuter:2005kb,Hind-Litim_Rahme}. The covariant form of this identification, $k^2 \sim R$, has been also a popular choice employed in studying the RG-improvement of gravitational actions in a cosmological context in \cite{Bonanno:2012jy, Hindmarsh:2012rc, Bonanno:2015fga,Copeland:2013vva}. Let us remind ourselves that the asymptotic expansion (\ref{HK-exp-Delta}) applies for sufficiently small curvature scales with $\bar{R}/k^2 < 1$. This fact, together with the slow--roll estimate $\bar{R}/ \Mp^2 \sim \lambda /\xi^2$ suggests the bound
\be
\left. \frac{k^2}{ \Mp^2} \right|_{\text{inflation} } \gtrsim \frac{\lambda}{\xi^2}. \label{k-infl}
\ee
This in turn places a bound on the value of $\tilde{\phi }_{*}$ assuming $\phi_{*} \sim \Mp/\sqrt{\xi}$,
\be
\tilde{\phi }_{* \text{inflation} } \equiv \left. \frac{ \phi_{* }}{k} \right|_{\text{inflation} }  \lesssim  \sqrt{\frac{\xi}{\lambda}}.
\ee
Given the above estimates, for the dimensionless product $G \phi_*^2$ which appears in the beta functions at non--zero v.e.v, it follows that
\be
\left. G \phi_*^2 = \widetilde{G} \tilde{\phi}_*^2 \right|_{\text{inflation} }  \sim \frac{1}{\xi}, \label{Gphi}
\ee
where we assumed that $G \simeq \Mp^2$ at energies well below the Planck mass. 

We can now get an estimate of the different terms in the equations (\ref{beta-G-Landau-vev})--(\ref{beta-lambda-Landau-vev}). We remind that the explicit expressions are given in (\ref{Appendix:beta-G-1L})--(\ref{Appendix:beta-lambda-1L}) of the Appendix. As an overall remark, notice the appearance of powers of $\xi \tilde{\phi}_*$ in the respective numerators, which can in principle acquire large values. Let us start with the quantity $\Omega$ which appears in the denominators and is defined in (\ref{Def:OMega-mu}). In the regime $\tilde{\phi}_*, \xi \gg1$ it can be approximated as
\be
\Omega \simeq 72 \pi \xi^2 G \phi_*^2 \sim 72 \pi \xi, \label{Omega-estimate}
\ee
where we used (\ref{Gphi}). 
We now look at the beta function for $\widetilde{G}$. Evaluating its denominator using (\ref{Omega-estimate}), it turns out it is of the order $\sim 10^4 \xi^2$. Its numerator consists of three different terms apart from the canonical one, a quadratic, cubic and quartic term in $\widetilde{G}$ respectively. In view of (\ref{Omega-estimate}) one finds for the overall coefficient of each of them in orders of magnitude that, $\sim 10^{-7} \cdot \widetilde{G}^2$, $\sim 10^{-6} \cdot \widetilde{G}^3$ and $\sim 10^{-3} \cdot \widetilde{G}^4$ respectively. Since $\widetilde{G} \ll 1$, the beta function will be dominated by the canonical term $= -2\widetilde{G}$ leading to 
\be
\left.  \widetilde{G}\right|_{\text{inflation} } \simeq \frac{k^2}{\Mp^2} \gtrsim  \frac{\lambda }{\xi^2} \ll 1, \label{G-inflation}
\ee
where we eliminated the arbitrary renormalisation scale $k_0$ by using the measured value $G = 1/\Mp^2$, and also used (\ref{k-infl}). 
Therefore, $G$ becomes constant and equal to its classical value. In Section \ref{sec:Higgs-intro}, most of the standard inflationary relations in the Einstein frame where modified by the quantity $x(k) \equiv G(k)/G_0$. Above result implies that $x(k) \simeq 1$, recovering the standard classical inflationary equations. 

The RG equations for $\xi$ and $\lambda$ (\ref{Appendix:beta-xi-1L}) --(\ref{Appendix:beta-lambda-1L}) also assume a non--trivial form. 
From above considerations, the denominator of $\beta_{\xi}$ is of the order $\sim 10^{10} \xi^3$, while the linear term in $\widetilde{G}$ in its numerator picks up a very large coefficient of the order $\sim 10^{4} \cdot \xi^4 \widetilde{G}$, however, when the latter is combined with the denominator it yields the overall estimate of $\sim 10^{-6} \cdot \xi \widetilde{G} \sim 10^{-6} \lambda/\xi$, using (\ref{G-inflation}). In a similar way of thinking, for the second-order term in $\widetilde{G}$ one can estimate $\sim 10^{-4} \cdot \xi^2 \widetilde{G}^2 \tilde{\phi}_*^2$, which in view of (\ref{Gphi}) and (\ref{G-inflation}) makes it of the order $\sim 10^{-4} \lambda/\xi$, while for the cubic term in $\widetilde{G}$ it turns out that it is of similar order, $\sim 10^{-3} \lambda/\xi$. 
In $\beta_{\lambda}$, the second and higher-order terms in $\widetilde{G}$ in its numerator appear coupled to large powers of $\xi$, e.g $\sim \xi^5 \widetilde{G}^2$  (\ref{G-inflation}). When the suppression coming from the denominator is taken into account, the quadratic term yields $\sim10^{-5} \xi^2 \widetilde{G}^2 \sim 10^{-5} \lambda^2/\xi^2$, and similar estimates result for the rest of the corresponding $\widetilde{G}$-terms in $\beta_{\lambda}$. As regards the running of the amplitude $\lambda/\xi^2$, using (\ref{beta-omega-Landau-vev}), one can see that it will also receive a suppression which will be at least of the order $\sim \lambda^2/\xi^4$. 

To summarise, the RG equations for a non-trivial v.e.v acquire a very involved, non--trivial form. The threshold effects from a sufficiently large v.e.v of the scalar in combination with the sufficiently low value of the cut--off $k$, act so as the terms from the gravitational sector in the RG equations receive a suppression in the sense discussed above. Above analysis also indicated a lower bound for the infrared, sliding RG scale $k$, presented in (\ref{k-infl}). A more precise estimate would require a detailed study of the field's dynamics and structure of the effective potential, which we will not pursue here. In the next section we will discuss the RG dynamics for the other limiting case, where $\tilde{\phi}_* \ll 1$.



\section{The post-inflationary regime} \label{sec:QDynamics-higher-energies}
We are now interested in the regime where $\tilde{\phi}_*$ is sufficiently small,
\be
 \tilde{\phi}_* \ll 1.
\ee
This occurs whenever the scalar has rolled down to a lower v.e.v $\phi_*$ with respect to some fixed energy scale (e.g after inflation), or as the cut--off scale $k$ increases while $\phi_*$ remains sufficiently small.
In the limit $\phi_*, v \to 0$ the exact beta function for $G$ acquires a simple form. From the flow equation (\ref{Flow:Effective-action}) it follows,
\begin{align}
 \beta_{\widetilde{G}} = (2 + \eta_{G}) \cdot \widetilde{G}, \; \; \; \; \;  \eta_{G} = \frac{14 \xi - 55} {2 \left(12 \pi -\tilde{G} \right) } \cdot \tilde{G}. \label{beta-G-Exact}
\end{align}
Notice that for $\widetilde{G} \lesssim 1$, $\xi \gg 1$, the anomalous dimension $\eta_G$ acquires a large and positive value, signalling a potentially singular behaviour in the running of $G$, however this is harmless for sufficiently low cut--off scales.
If we expand $\eta_G$ for $\widetilde{G} \ll 1$ to linear order, we arrive at the previously found 1-loop equation but with $\tilde{\phi}_* \to 0$,
\begin{align}
\left. \beta_{\widetilde{G}} \right|_{\text{1--loop}} =2 \tilde{G} 
+c \cdot  \tilde{G}^2 + \mathcal{O}(\tilde{G}^3), \; \; \; \; \;  c \equiv \frac{1 }{24 \pi }\left(14 \xi -55\right)
 \label{beta-G-Landau}.
\end{align}
It exhibits two fixed points for $\widetilde{G}$, the trivial (Gaussian) one with $\widetilde{G} = 0$, and a non--trivial fixed point at 
\be
\widetilde{G}_{\text{fp}} = \frac{48 \pi}{55 - 14 \xi}, \label{fp-G-approx}
\ee
which becomes negative for $\xi > 55/14$, and it is always attractive.

For the rest of the equations (\ref{beta-xi-Landau-vev})--(\ref{beta-lambda-Landau-vev}), when evaluated in the limit $\tilde{\phi}_*  \to 0$, all the non-standard terms $\sim \tilde{\phi}_*^2 \widetilde{G}$ vanish. The exact equation for $\xi$ in this limit is given in (\ref{xi-exact-Appendix}). 
Since we are below the Planck mass, it is enough for our purpose to present the respective equations under the 1-loop approximation
%
%
\begin{align}
\left.  \beta_{\xi} \right|_{\text{1--loop}} = \frac{\lambda  (28 \xi +5)}{64 \pi ^2}
+ \frac{\xi ^2 (48 \xi +31)}{8 \pi } \cdot  \tilde{G}
%
 \label{beta-xi-Landau},
\end{align}
\begin{align}
& \left. \beta_{\lambda} \right|_{\text{1--loop}}= 
 \frac{21 \lambda ^2}{16 \pi ^2} 
+ \frac{\lambda}{\pi}  \left(36 \xi ^2+14 \xi +5\right) \cdot \tilde{G}  
\label{beta-lambda-Landau}.
\end{align}
%
Note that the gravitational corrections enter with a positive sign. In view of (\ref{G-inflation}), the terms $\sim \xi^3 \widetilde{G}$ and $\sim \lambda \xi^2 \widetilde{G}$ in above equations are of order $\lambda \xi$ and $\lambda^2$ at the scale given by (\ref{k-infl}). As the cut--off is decreased though, they tend to decrease sufficiently fast as $\widetilde{G} \lll 1$.
We can derive approximate, analytic solutions for equations (\ref{beta-G-Landau})-(\ref{beta-lambda-Landau}) in the regime where $\xi \gg 1$, and assuming that initially $\widetilde{G}$ is sufficiently small, so that the equations are dominated by the standard terms. This allows us to set $\widetilde{G} = 0$ in (\ref{beta-xi-Landau}) and (\ref{beta-lambda-Landau}). 
Under these assumptions, and for $\xi \gg1$, (\ref{beta-xi-Landau}) and (\ref{beta-lambda-Landau}) yield the familiar solutions \footnote{Notice that the exponent of the denominator in the solution for $\xi$ depends on the regularisation scheme used.}
\be
\lambda(k) \simeq \frac{\lambda_0}{1 - \frac{21 \lambda_0}{16 \pi^2} \log (k/k_0)}, \; \; \; \; \; \; \;   \xi(k) \simeq \frac{\xi_0}{\left( 1 -  \frac{21 \lambda_0}{16 \pi^2} \log (k/k_0) \right)^{1/3}}, \label{Solutions-lambda-xi}
\ee
with $\lambda_0 = \lambda(k = k_0)$ and $\xi_0 = \xi(k = k_0)$, and $k_0$ an arbitrary energy scale.

Now, looking at equation (\ref{beta-G-Landau}) one can see that for $\xi \gg 1$, the coefficient of the leading-order correction in $\widetilde{G}$ is $c \simeq 7 \xi/(12 \pi) > 0$. 
If we neglect the logarithmic running of $\xi$ and assume that $\xi \simeq \xi_0 \equiv  \text{constant}$ in (\ref{beta-G-Landau}), then for $c \simeq 7 \xi_0/(12 \pi)$ we can solve (\ref{beta-G-Landau}) for $\widetilde{G}(k)$,
\be
\widetilde{G}(k) \simeq \frac{ (k^2/\Mp^2)}{1 - (\nicefrac{7}{24\pi}) \cdot \xi_0 \cdot (k^2/\Mp^2) }. \label{Solution-G}
\ee
We have traded the arbitrary constants in the solution for the renormalised values $\widetilde{G}_R = \widetilde{G}(k = k_R)$ with $k_R/k_0 \ll 1$. For $k^2/\Mp \ll 1$ the solution (\ref{Solution-G}) decreases quadratically, entering deep into the classical regime with $G = 1/\Mp^2$.
Solution (\ref{Solution-G}) suggests that as we raise the cut--off from smaller to higher values, its denominator becomes zero at 
\be
k = \sqrt{\frac{24 \pi}{7}} \cdot \frac{\Mp}{\sqrt{\xi_0}}. \label{k-pole}
\ee
Of course, by the moment $\widetilde G \simeq  1$, the approximate solutions (\ref{Solutions-lambda-xi}), (\ref{Solution-G}) are not valid anymore. The above diverging behaviour is unphysical and cannot exist in reality, as it would suggest that gravity becomes strongly coupled at a scale below the Planck mass. Most importantly, the scale defined through (\ref{k-pole}) is beyond the lower bound on the inflationary cut--off scale, given in (\ref{k-infl}), which implies that by that moment the scalar should have acquired a sufficiently large v.e.v, preventing the coupling to hit the pole. Notice that the scale (\ref{k-pole}) coincides with the lower value of $\phi_*$ during inflation, implying that at the scale (\ref{k-pole}) $\phi_*/k \sim \mathcal{O}(1)$; beyond this scale, $\tilde{\phi}$ could become sufficiently small, turning off the suppression of gravitational effects as described earlier. In this sense, (\ref{k-infl}) also provides an extreme upper bound for the infrared RG scale at inflation $k \sim k_{\text{inflation}}$.

\section{Asymptotic Safety} \label{sec:Asymptotic-safety}
It is interesting to briefly discuss the possible embedding of the Higgs inflationary action within the scenario of Asymptotic Safety (AS). 
Within AS \cite{Weinberg:1980gg} \footnote{For reviews see Refs. \cite{Morris:1998da, Gies:2006wv, Niedermaier:2006wt,  Percacci:2007sz, Percacci:2011fr, Litim:2008tt, Reuter:2012id, Reuter:2012xf}.}, the UV completion of a theory is achieved under the existence of a UV fixed point under the RG. 
For gravity, there are growing indications that this is the case in different setups ranging from Einstein--Hilbert \cite{Reuter:1996cp, Falkenberg:1996bq, Souma:1999at, Lauscher:2001ya,Lauscher:2002sq, Litim:2003vp, Reuter:2001ag, Dou:1997fg,Falls:2014zba,Eichhorn:2013ug,Vacca:2010mj} and higher-derivative gravity \cite{Granda:1998wn, Codello:2006in, Codello:2007bd,  Codello:2008vh, Machado:2007ea, Benedetti:2009gn,Benedetti:2009rx, Benedetti:2012dx, Benedetti:2011ct, Falls:2013bv, Manrique:2011jc,Demmel:2013myx, Dietz:2012ic, Dietz:2013sba, Falls:2014tra, Benedetti:2013jk, Demmel:2015oqa,Ohta:2015efa}, to scalar--tensor \cite{Narain:2009fy,Narain:2009gb, Percacci:2015wwa} and unimodular gravity as examples \cite{Eichhorn:2013xr, Saltas:2014cta, Eichhorn:2015bna,Benedetti:2015zsw}. The cosmological consequences and phenomenology of the scenario have been studied in various works \cite{Wein, Reuter:2005kb,Hindmarsh:2012rc,Contillo:2011ag,Cai:2011kd,Hindmarsh:2011hx, Xianyu:2014eba, Bonanno:2001xi,Reuter:2005kb, Shaposhnikov:2009pv, Reuter:2012xf, Copeland:2013vva,Grozdanov:2015zna}. 

Within the context of Higgs inflation, AS could provide a fundamental framework towards a UV completion, and more solid ground for the behaviour of quantum gravitational corrections at very high energies. 
To calculate the fixed-point structure of the action (\ref{Higgs-action}) one needs the full, non--perturbative set of beta functions, which correspond to the solutions of the flow equation (\ref{Flow:general}) with the potentials $f$ and $V$ given by (\ref{ansatz-f}) and (\ref{potential-unbroken}) below. In this section we will set $r=1$ for the regulator parameter, and will work in the deep UV regime where $k \to \infty$, $\phi_{*}/k \to 0$. In this regime, we expand around $\phi_* = 0$ as 
\be
V(\phi) = u + m^2 \phi^2 + \lambda \phi^4. \label{potential-unbroken}
\ee
Notice that the couplings in (\ref{potential-unbroken}) are related to the ones in (\ref{potential-broken}) through $u =  v + (1/4)\lambda \phi_{*}^4$, $m^2 = - \lambda \phi_{*}^2$.
The zeros of the full system of beta functions extracted from (\ref{Flow:Effective-action}) suggest the theory possesses one UV fixed point (UVFP), where all scalar-field interactions become trivial, while Newton's constant and vacuum energy, $\widetilde{G}$ and $\widetilde{v}$, are interacting,
\begin{align}
\widetilde{G}_{\text{fp}} = 0.527, \; \; \widetilde{u}_{\text{fp}}  = 0.019, \; \; \xi_{\text{fp}}  = 0, \; \; \tilde{m}_{\text{fp}}^2 = 0, \; \; \lambda_{\text{fp}}  = 0. \label{UVFP}
\end{align}
This is the well--known Gaussian-matter fixed point (GMFP), due to the vanishing of the matter interactions, and has been previously studied in a similar setup \cite{Narain:2009fy, Narain:2009gb}. The attractivity properties in the vicinity of the UV fixed point reveal the relevance/irrelevance of the couplings in the UV. To find out, we calculate the linearised RG flow around (\ref{UVFP}) from which we can straightforwardly extract the associated eigenvalues. In particular, it turns out that
\begin{align}
\lambda_{\tilde u, \widetilde G} = -0.243 \pm 4.024i, \; \; \; 
\lambda_{\xi, \tilde{m}^2}= -2.43 \pm 4.024i, \; \; \; 
 \lambda_{\lambda} = 4.462. \label{UVFP-Eigenvalues}
\end{align}
The eigenvalues associated with the vacuum energy and Newton's coupling form a complex conjugate pair, with a negative (attractive) real part, and a similar situation occurs for $\xi$ and $m^2$. Notice that the non--minimal coupling $\xi$ is relevant, while the quartic one, $\lambda$ (marginal in power counting), becomes irrelevant. 
The connection with AS would require that the initial conditions along the RG flow set at the end of inflation reach the UV fixed point when evolved under the RG flow. 
Sufficiently close to the fixed point, for the stability of Newton's coupling, according to the discussion around \ref{beta-G-Landau2} and from equation (\ref{beta-G-Exact}) respectively, we see that a necessary condition is \footnote{Notice however that the exact value of the coefficient depends on the regulator choice. More details around this are discussed in Section \ref{sec:Gauge-regulator-etc}.}
\be
\xi< \frac{55}{14} \simeq 3.93. \label{bound-xi}
\ee
Of course, the pole (\ref{k-pole}) should be also avoided in evolving from sufficiently low scales, but this is what one would expect to happen taking into account the running of the v.e.v $\phi_*$, remembering that (\ref{k-pole}) corresponds to the vacuum case. A study of this issue would require a detailed numerical study of the complete set of beta functions, which we leave for a future work. 

\begin{figure}
\begin{center}
\includegraphics[scale=0.63]{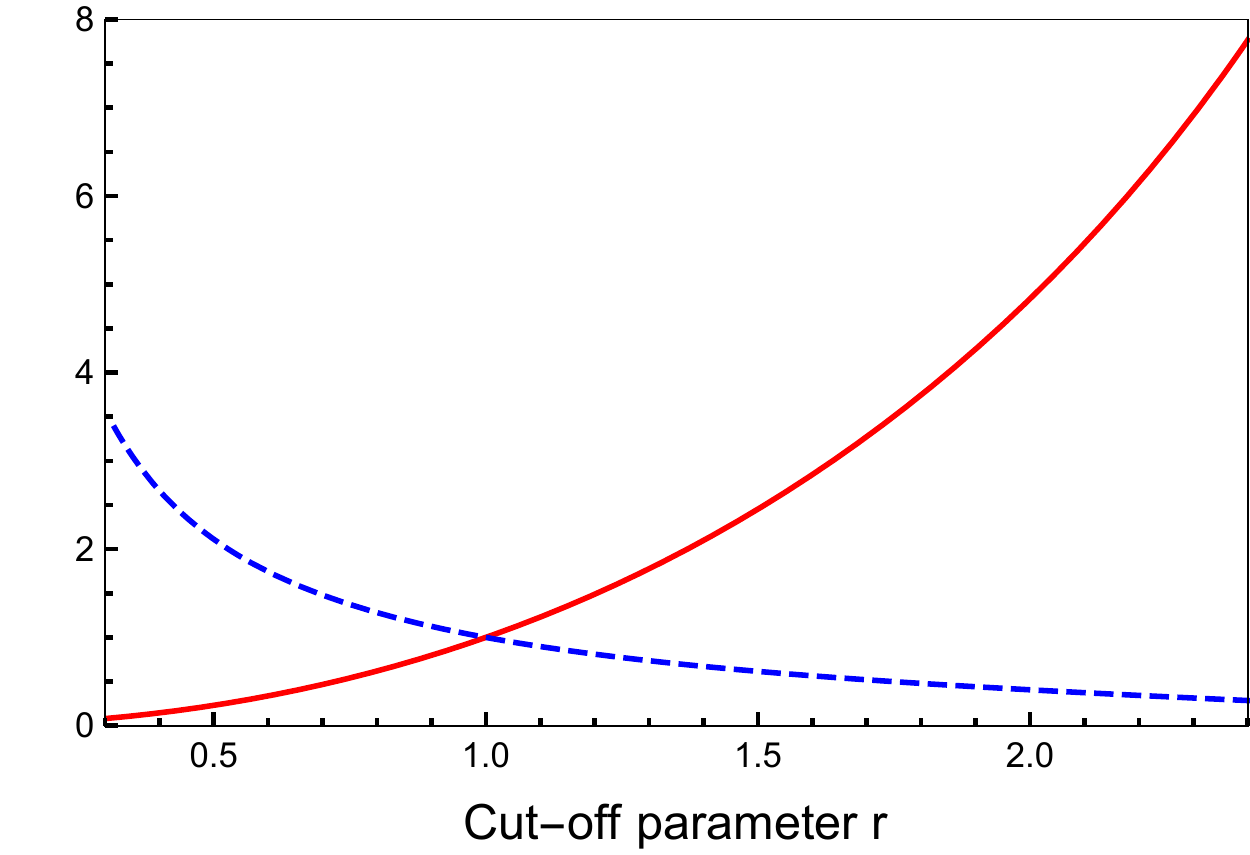}
\caption{The figure illustrates the variation of the non-trivial fixed-point values in (\ref{UVFP}) with the regulator parameter $r$ (see also Section \ref{sec:Quantum-calculation}) with respect to its optimal value, i.e $\tilde{u}_{\text{fp}}(r)/\tilde{u}_{\text{fp}}(r=1)$ (Blue, dashed) and $\widetilde G_{\text{fp}}(r)/\widetilde{G}_{\text{fp}}(r =1)$ (Red, continuous). }
\label{fig:Reg-Depend}
\end{center}
\end{figure}


\section{A comment on the gauge and regulator dependence} \label{sec:Gauge-regulator-etc}

The use of a truncated theory space in combination with working off--shell introduces a dependence on the regulator and gauge choice respectively \footnote{For issues regarding gauge and regulator dependence within the functional RG see for example \cite{Morris:1998kz, Arnone:2006ie,Falkenberg:1996bq,Narain:2009qa,Pawlowski:2015mlf}.}. 
The explicit dependence on the regulator parameter $r$ significantly increases the complexity of the equations, so we only explicitly discuss its influence on the renormalisation of $G$ at leading order and on the UV fixed-point values respectively, for the vacuum case. The same is true for the gauge parameter, and below we will briefly discuss the case of the Landau gauge.

The leading order correction in the equation for the renormalisation of $G$, equation (\ref{beta-G-Landau}), has been crucial for the earlier analysis. With an unspecified regulator parameter ($r$) it reads
\be
\beta_{\widetilde{G}} \simeq 2 \widetilde{G} +  c\cdot \widetilde{G}^2 + \mathcal{O}(\widetilde{G}^3), \; \; \; c = \frac{1}{24\pi}\left( -28 +\frac{45}{r}- \frac{72}{r^2} + \frac{14 \xi }{ r^2}\right). \label{beta-G-reg-gauge}
\ee
The $\xi$-independent terms in $c$ give a negative contribution for all $r>0$, while for $\xi \gg 1$, $r$ would also have to be also very large to make the contribution of $\xi$ unnoticeable. One is here reminded that, $r =1$ corresponds to the ''optimal" value of the regulator function \cite{Litim:2000ci} (see Section \ref{sec:Quantum-calculation}), and one should not expect large deviations from it. It is also interesting to notice that the UV fixed point exists as long as  $r \sim \mathcal{O}(1)$, in particular $0.33 \lesssim r  \lesssim 3.6$ (see also Figure \ref{fig:Reg-Depend}).

The beta functions presented here have been also calculated in \cite{Narain:2009fy} using a different field decomposition and in the {\it Landau gauge ($\alpha = 0$)}. Let us write here the results found there for $G$ at 1--loop,
\be
\beta_{\widetilde{G}} \simeq 2\widetilde{G} + \frac{1}{24 \pi}(24 \xi - 77)\cdot \widetilde{G}^2, \label{beta-G-Percacci}
\ee
and we have performed a similar check for the beta functions for $\xi$ and $\lambda$. Notice that the order of magnitude and signs of the coefficients in (\ref{beta-G-Percacci}) are in agreement with the ones presented here.

\section{Summary}\label{sec:Discussion}
We employed the functional RG to study quantum corrections for the Higgs inflationary action during inflation and beyond, including the effect of gravitons. The formalism employs the Wilsonian approach to the RG, which provides an effective description of the physical system from small to large scales, as the infrared RG scale is moved in a continuous way. What is more, its extension to the non--perturbative realm allows for a connection with the Asymptotic Safety scenario for quantum gravity. 
Within this framework, we evaluated the exact RG flow for the Higgs-gravity effective action, and explicitly studied the resulting RG equations including the leading-order gravitational corrections at 1--loop, under the particular assumptions described in Section \ref{sec:Quantum-calculation-Setup} (see Appendices \ref{Appendix:Flow} and \ref{sec:Betas-approx} for explicit expressions). In particular, the calculation was performed in the Jordan frame and for the background of a Euclidean de Sitter. 
%

During inflation, the corrections coming from the gravitational sector acquired a non-trivial form, with the new terms generated under the RG due to the scalar's non--zero v.e.v $\phi_* \sim \Mp/\sqrt{\xi}$, introducing appropriate threshold effects which allowed for a suppression to the running of the couplings such as the quartic interaction $\lambda$, non--minimal coupling $\xi$ and Newton's coupling, in the sense explained in Section \ref{sec:Q-dynamics-inflation}. In particular, in this regime, Newton's $G$ presented a negligible running, reducing to its constant, classical value. The sliding RG scale $k$ within this framework is interpreted as the typical energy or coarse-graining scale of the system. The consistency of the approach placed a lower bound on its value during inflation, suggesting it to be of the order $\sim \sqrt{\lambda} \Mp/\xi$ (see the discussion in Section \ref{sec:Q-dynamics-inflation}), which lies a few orders of magnitude below the Planck scale. As long as the v.e.v of the scalar dropped to sufficiently low values after inflation, with gravity entering the deep classical regime at lower cut--off scales, the RG equations acquired their standard perturbative form, allowing for a connection with the low-energy regime.  
What is more, as discussed in Section \ref{sec:Asymptotic-safety}, at arbitrarily high energies, the theory possesses the well known Gaussian-matter UV fixed point, which could provide a connection of the model with the scenario of Asymptotic Safety. In particular, the RG dynamics would be expected to drive the large initial value for $\xi$ to smaller values at higher energies, eventually reaching its fixed-point at $\xi = 0$. 

To conclude, Higgs inflation could provide with a promising framework for the early universe and a natural extension of the standard model of particle physics. The investigation of its connection with the physics of higher energies and a potential UV completion, including gravity, are natural questions to ask. The issue of the possible instability of the Higgs potential due to the influence of gauge/Yukawa couplings has not been considered in this work, and its study poses a challenging issue within this context. What is more, in view of our original action ansatz, the higher-order operators generated under the RG procedure were neglected, and their study could provide further insights about the model, such as the issue of unitarity violation. An analysis of the structure of the RG dynamics beyond 1--loop and the connection with Asymptotic Safety is yet another challenging task. From the discussion of Sections \ref{sec:QDynamics-higher-energies} and \ref{sec:Asymptotic-safety} it turns out that in this direction, a consistent study of the full system of RG equations taking into account the running of the scalar's v.e.v is required. We hope that this work will motivate further studies of the model within the functional RG and/or Asymptotic Safety. 



\appendix

\section{Evaluation of the ERGE} \label{Appendix:Expl-rel}

Here we present more explicit steps for the calculation of Section \ref{sec:Quantum-calculation}.
Our starting point is the action
\be
\Gamma[g_{\mu \nu}, \phi]  = \int \sqrt{g} \left[ - f(\phi)R + \frac{1}{2}g^{\mu \nu} \partial_\mu \phi \partial_\nu \phi  + V(\phi) \right].  \label{Action-ansatz}
\ee
Under the field expansion of the metric and scalar field around a constant background ($\bar{g}_{\mu \nu}, \bar \phi$), as
$g_{\mu \nu} =   \bar{g}_{\mu \nu} + h_{\mu \nu}, \; \phi = \bar{\phi} + \delta \phi$ we expand up to second-order as
\begin{align}
& \delta^{(2)} \Gamma  \equiv \int \sqrt{g} \, \Phi_A  \cdot \Gamma^{AB} \cdot \Phi_B = \frac{1}{2} \int \sqrt{g} f(\phi) \Big[ - h_{\rho \nu} h^{\rho \sigma} R^{\nu}{}_{\sigma} 
+  h h^{\rho \sigma}R_{\rho \sigma}    +    \frac{1}{2}h_{\rho \sigma} h^{\rho \sigma} R     -     \frac{1}{4}h^2 R    +   R_{\rho}{}^{\mu \nu}{}_{\sigma}h_{\mu \nu}h^{\rho \sigma}   \no \\
& - \frac{1}{2}h_{\rho \sigma}\Box h^{\rho \sigma}     + \frac{1}{2}h \Box h   -     \nabla^{\kappa} h_{\kappa \mu} \nabla^{\lambda} h_{\lambda}{}^{\mu}    +    \nabla^{\kappa}h_{\mu \kappa} \nabla^{\mu} h 
+ V(\phi) (\frac{1}{2}h^2 - h_{\alpha \beta}h^{\alpha \beta}) \Big] \no \\
& -   2f'(\phi) \left[ \frac{1}{2} R h \cdot \delta \phi + \delta \phi \cdot (-h^{\alpha \beta}R_{\alpha \beta} - \Box h + \nabla_{\alpha} \nabla_{\beta} h^{\alpha \beta})\right] + V'(\phi) h \delta \phi \no \\
& +   \delta \phi \Big[  V''(\phi) - f''(\phi) R   + (-\Box) \Big]  \delta \phi . \label{Eff-action-Sec-order}
\end{align}
Notice that derivatives and curvature tensors in (\ref{Eff-action-Sec-order}) are built out of $\bar{g}_{\mu \nu}$.
From (\ref{Eff-action-Sec-order}) it is a straightforward excercise to extract the individual entries of $\Gamma^{A}_{B}$, corresponding to the different vertices. They read as
\begin{align}
\Gamma_{h_{\mu \nu} \cdot h_{\mu \nu}}^{\alpha \beta \gamma \delta} =   \frac{f(\phi)}{2}\cdot & \Bigg[ 
    \frac{1}{2}\left(  g^{\gamma (\alpha}g^{\beta) \delta} +  g^{\alpha (\gamma}g^{\delta) \beta}   -g^{\alpha \beta} g^{\gamma \delta} \right) (-\Box)  
    +  \frac{1}{2} \left( g^{\gamma (\beta} \delta^{\alpha)}_{\kappa}  \delta^{\delta}_{\lambda} +  g^{\beta (\gamma} \delta^{\delta)}_{\lambda}  \delta^{\alpha}_{\kappa} - 2 g^{\alpha \beta} \delta^{(\gamma}_{\kappa} \delta^{\delta)}_{\lambda}  \right) \nabla^{\kappa} \nabla^{\lambda} \no \\
%
& + \frac{1}{2} \left( g^{\alpha \beta}g^{\gamma \delta} -  g^{\gamma (\alpha}g^{\beta) \delta}  -  g^{\alpha (\gamma}g^{\delta) \beta}   \right)V(\phi)
+ 2R^{\gamma \alpha \beta \delta}
- \left(  g^{\gamma (\alpha} R^{\beta) \delta} + g^{\alpha ( \gamma } R^{ \delta) \beta}   - 2g^{\alpha \beta}R^{\beta \gamma}    \right) \no \\
& 
+ \left(  g^{\gamma (\alpha} g^{\beta) \delta} + g^{\alpha ( \gamma } g^{ \delta) \beta}   - \frac{1}{2}g^{\alpha \beta} g^{\gamma \delta}  \right)R
  \Bigg],
\end{align}
\begin{align}
\Gamma_{h_{\mu \nu} \cdot \phi}^{\alpha \beta} &= f'(\phi) \cdot \left[ \left( \delta^{(\alpha}_{\mu}  \delta^{\beta)}_{\nu} + g^{\alpha \beta} g_{\mu \nu}\right)\nabla^{\mu} \nabla^{\nu}  - (R^{\alpha \beta} - \frac{1}{2} g^{\alpha \beta}R)  - g^{\alpha \beta}\frac{V'(\phi)}{2 f'(\phi)} \right],
\end{align}
\begin{align}
\Gamma_{\phi \cdot \phi}^{\alpha \beta} &= -(-\Box) + f''(\phi)R - V''(\phi).
\end{align}

Similarly as before, for the gauge--fixing and ghost operators respectively we have
\begin{align}
&S_{GF}^{\alpha \beta \gamma \delta} = - \frac{f(\phi)}{2\alpha } \left[ 
\frac{1}{2} \left(   g^{\delta (\alpha}\delta^{\beta)}_{\kappa}\delta^{\gamma}_{\lambda} 
+ g^{\beta (\gamma}\delta^{\delta)}_{\lambda}\delta^{\alpha}_{\kappa} -  2g^{\gamma \delta}\delta^{(\alpha}_{\kappa}\delta^{\beta)}_{\lambda}\right)  \nabla^{\kappa} \nabla^{\lambda}
- \frac{1}{4} g^{\alpha \beta} g^{\gamma \delta} (-\Box) \right],  \\
&S_{\text{ghost}}{}^{\mu}{}_{\nu} = - \left(  - \Box  - \frac{1}{4}R \right)\delta^{\mu}{}_{\nu} .
\end{align}
Under the trace expansion (\ref{Field-expansion}) and the background choice of a Euclidean sphere where $R_{\alpha \beta \gamma \delta} = (R/12) \cdot (g_{\alpha \gamma} g_{\beta \delta} - g_{\beta \gamma} g_{\alpha \delta})$, the different inverse propagator entries take a simpler form which schematically reads as
\be
\Gamma_{\Phi_A \cdot \Phi_B} =  Z_{\Phi_A \Phi_B}(\phi, R; r)\cdot \left( -\Box \right)  +  U_{\Phi_A \Phi_B}(\phi, R; r).
\ee
The regulators which will serve as to cut--off the eigenvalues of the Laplacian which' value is less than the infrared cut--off $k$ are appropriately defined as
$
R_{ k \; \Phi_A \Phi_B} \equiv   Z_{\Phi_A \Phi_B} \cdot R_k \left( -\Box; r\right).
$
Under the modification of the Hessians, $\Gamma_{ k \; \Phi_A \Phi_B}(- \Box) \to \Gamma_{ k \; \Phi_A \Phi_B}(- \Box) + R_{ k \; \Phi_A \Phi_B}(- \Box)$, the regulators $R_{ k}$ will combine with the associated laplacian operators, which corresponds to the choice of a Type 1a cut--off. 
With above relations and definitions, the calculation of the trace integral in the ERGE (\ref{ERGE}) reduces to the evaluation of the trace over momenta 
\be
 \frac{1}{2} \text{Tr} \left[ \left( \Gamma_{\Phi_A \Phi_B} + R_{\Phi_A \Phi_B}  \right)^{-1} \cdot \partial_{t} R_{\Phi_A \Phi_B} \right],
\ee
where it is understood that the first term stands for a matrix inverse, and the dot corresponds to a matrix multiplication respectively. 
Defining $P_k(-\Box) \equiv -\Box + R_k(-\Box)$, the trace can be evaluated as \cite{Reuter:1996cp,Dou:1997fg,Cod-Perc-Rahme1}
\begin{align}
 \text{Tr} & \left[ \frac{g(-\Box) }{P_{k}(-\Box) + { U(\bar{R})}} \right]  =   \frac{1}{(4\pi)^{2}} \sum_{i}^{\infty} \sum_{l = 0}^{\infty} Q_{2-i}\left(\frac{ g(-\Box) }{P_{k}^{l+1}(-\Box)} \right) \cdot (-1)^{l}  \int d^{4}x\sqrt{g} \, \text{tr}({ U})^{l} \, a_{2i}(-\Box), \label{HK-expansion}
\end{align}
with the definition of the functionals $Q_{2-i}$
\be
Q_{2-i}\left(\frac{ g(z)}{P_{k}^{l+1}(z)} \right)  = \int_{0}^{\infty}ds e^{-zs}\tilde{g}(z),
\ee 
ands $z \equiv -\Box$. The function $g$ denotes either $g \equiv R_{k}$ or $g \equiv k\partial_k R_k$, while $\tilde{g}$ stands for the anti-Laplace transform of $g$. $a_{2i}(-\Box)$ correspond to the heat kernel coefficients of the Laplacian. For more explicit details we refer to Refs \cite{Reuter:1996cp,Dou:1997fg,Reuter:2012id,Reuter:2007rv,Lauscher:2001ya,Cod-Perc-Rahme1}.


\subsection{The flow of the effective action} \label{Appendix:Flow}
The flow of the effective action according to the ERGE organises itself in the following way,
\begin{align}
\partial_t \Gamma_k = \mathcal{F}_0 + \mathcal{F}_{1} \cdot  \frac{\partial_t f}{f} + \mathcal{F}_{2}\cdot \frac{\partial_t f'}{f'}, \label{Flow:Effective-action}
\end{align}
with primes here denoting derivatives with respect to $\tilde{\phi} \equiv \phi/k$ and $\partial_t \equiv k \partial_k$.
It is convenient to work with the dimensionless quantities measured in units of the cut--off $k$, 
\be
\tilde{f} \equiv \tilde{f} (\tilde{\phi}) \equiv f/k^2, \; \; \; \widetilde{V} \equiv \widetilde{V}(\tilde{\phi})  \equiv V/k^4, \; \; \; \tilde{R} \equiv R/k^2.
\ee
Introducing the convenient quantities $\sigma \equiv \widetilde{V}/\tilde{f}$ and $\omega \equiv (3 + \tilde{R})\tilde{f}' + \widetilde{V}'$,
the individual terms appearing in the flow equation (\ref{Flow:Effective-action}) are defined as follows,

\begin{align}
\mathcal{F}_0 & \equiv - \frac{1}{4\pi^2} \left( 1 + \frac{7}{12} \tilde{R} \right) + \frac{1}{16 \pi^2} \cdot  \frac{9(3 + \tilde R)}{\mathcal{D}_0} 
 + \frac{1}{8 \pi^2} \cdot \frac{ \tilde{f}' \omega (3 + \tilde{R}) }{ \mathcal{D}_2}  \nonumber \\
%
%
& + \frac{1}{192 \pi^2} \cdot \frac{\tilde{f}}{\mathcal{D}_1} \Bigg[ 2\left( 14 + 5 \tilde{R}  \right) \left( \sigma - 1\right) 
-8(3 + \tilde{R})  \left( 1 - \tilde{R} \tilde{f}'' + \widetilde{V}'' \right) \Bigg], \nonumber
\end{align}
\begin{align}
& \mathcal{F}_{1} \equiv \frac{(2 + \tilde R) \tilde{f}}{64 \pi^2} \cdot \left[  \frac{9}{\mathcal{D}_0}   
-   \frac{2}{3} \cdot \frac{1 - \tilde{R} \tilde{f}'' - \widetilde{V}''}{\mathcal{D}_1} \right], 
%
%
\; \; \;  \; \; \mathcal{F}_{2} \equiv \frac{1}{32 \pi^2} \cdot  \frac{ (2 + \tilde{R} )\tilde{f}'  \omega}{ \mathcal{D}_1}, 
\end{align}
\begin{align}
& \mathcal{D}_0 \equiv \tilde{f} \left(3 + 2 \tilde{R} - 3\sigma \right) , \; \; \; \; 
 \mathcal{D}_1 \equiv \tilde{f}\left[ - \omega^2/\tilde{f} + 2\left(  \sigma -1 \right)\left( 1 - R \tilde{f}'' + \widetilde{V}'' \right) \right], \nonumber \\
& \mathcal{D}_2 \equiv \tilde{f}\left[   2(1  -  \sigma) + \omega^2/\tilde{f} + 2\left(  1 - \sigma \right) \left( \widetilde{V}'' -  \tilde{R} \tilde{f}'' \right)  \right], \label{F-explicit}
\end{align}
Equation (\ref{Flow:Effective-action}) describes the change of $\Gamma_k$ under an infinitesimal change of the RG scale $k$. As expected, the flow of the effective action depends only up to second derivatives with respect to the scalar $\phi$, and up to first-order derivatives with respect to the RG scale $k$. Notice the RG derivatives on the right-hand side which reflect the RG-improvement beyond the 1--loop level. A similar flow equation has been previously derived in \cite{Narain:2009fy} using a different field decomposition and evaluated in the Landau gauge.  

\subsection{Flow of the scalar potentials $V$ and $f$ at 1--loop} \label{Appendix:Flow-1loop}
It is instructive to evaluate the 1--loop approximation of the flow equation, which corresponds to switching off the RG derivatives on the right-hand side of (\ref{Flow:Effective-action}), see also (\ref{Gamma-1-loop}). In what follows, primes will denote derivatives with respect to $\tilde{\phi}$. Then, the flow of each potential is described by the following equations 
\begin{align}
 \partial_t  \widetilde{V} = \left( - 4 + \eta_{{V}}  \right)\widetilde{V}, \hspace{0.7cm}
\partial_t  \widetilde{f} =  \left(  - 2  + \eta_{{f}} \right) \tilde{f},
\end{align}
with the anomalous dimensions of the potentials $f$ and $V$ respectively taking the following form
\begin{align}
\eta_{{f}}           \equiv  \frac{\partial_t f}{f}          =    ( {\bf A}_{\tilde f} \cdot {\bf c}_{\tilde{f}} )^{T} \cdot {\bf d}_{\tilde{f}},  \hspace{0.7cm} 
\eta_{{V}}  \equiv \frac{\partial_t V}{V}   =   ( {\bf A}_{\tilde V} \cdot {\bf c}_{\tilde{V}} )^{T} \cdot {\bf d}_{\tilde{V}}.
\end{align}
Above matrices are defined as
\begin{align}
{\bf A}_{\widetilde{f}}  \equiv
\frac{1}{386\pi ^2} \cdot  \frac{1}{\mathcal{D}^2} \cdot \left(
\begin{array}{ccc} 
-55        &    2          & -19   \vspace{0.15cm}\\
14         &   -28        & 14    \vspace{0.15cm}\\
-115      &   14         & -43   \vspace{0.15cm}\\
-60        &   12         & -24 \\
\end{array}
\right)
, \; \; \; 
{\bf c}_{\widetilde{f}}  \equiv
 \left( 
\begin{array}{c} 
    1  \\
    \sigma   \\
    \sigma^2 \\
\end{array}
 \right), 
 \; \; \; 
{\bf d}_{\widetilde{f}}  \equiv
 \left(
 \begin{array}{c} 
 1 \\  
 \tilde{f}'' \\
 \sigma \tilde{f} \cdot (\widetilde{V}''/\widetilde{V}) \\  
 \sigma^2 \tilde{f}^2 \cdot (\widetilde{V}''/\widetilde{V})^{2} \\
\end{array}
\right),
\end{align}
\begin{align}
{\bf A}_{\tilde V}  \equiv
\frac{1}{192\pi ^2} \cdot  \frac{1}{\mathcal{D}}  \cdot  \left(
\begin{array}{cccc} 
86    \hspace{0.1cm}      &   -(41 +11 \sigma )     \hspace{0.1cm}       & -34     \vspace{0.15cm}\\
378                                                          &   108                                                                          & 0         \vspace{0.15cm}\\
432                                                          &   -216                                                                         & 0         \vspace{0.15cm}\\
120                                                          &    96                                                                           &  0        \vspace{0.15cm}\\
72    \hspace{0.1cm}     & -24                                  \hspace{0.1cm}        & -48      \\
\end{array}
\right)
, \; \; \; 
{\bf c}_{\tilde{V}} =  \left( 
\begin{array}{c} 
  1       \\
  \sigma   \\
    \sigma^2    \\
\end{array}
 \right),
 \; \; \; 
{\bf d}_{\tilde{V}}  \equiv
 \left( 
\begin{array}{c} 
 1 \\
\tilde{f}'^2/\tilde{f}   \\
\sigma \tilde{f}'  \cdot  (\widetilde{V}' / \widetilde{V})     \\
\sigma^2 \cdot ( \widetilde{V}' / \widetilde{V})^{2}      \\
\sigma \tilde{f} \cdot (\widetilde{V}''/\widetilde{V}) \\
\end{array}
 \right),
\end{align}
together with the definition of the quantity $\mathcal{D}$
\begin{align}
\mathcal{D} \equiv  \tilde{f}^2 \cdot ( 1 - \sigma) \left[ 2\left(1 -  \sigma \right) \left(1 + \widetilde{V}'' \right)  + \frac{\left( 3 \tilde{f}' - 2\widetilde{V}'\right)^2}{\tilde{f}} 
\right]. 
\end{align}


\section{Explicit relations for the running couplings} \label{sec:Betas-approx}

Here we provide the {\it 1-loop expressions} used in the text at non-zero v.e.v for $\phi$, and fixing the regulator parameter $r = 1$ for simplicity. They are extracted from (\ref{Flow:Effective-action}), first expanding the potential as in (\ref{potential-broken}) (neglecting $v_k$), and then projecting out the corresponding operators appearing on each side of the equation. Defining $Z_G \equiv 1/(16 \pi G)$, we have


{\small
\begin{align}
\left. -  \left( \frac{\partial_t Z_G}{Z_G}\right) \right|_{\text{1-loop}} & =  (16 \pi \widetilde{G}) \cdot \Bigg[ 14 \xi  -5 \left(48 \mu^2 + 46 \mu +11\right)
+ 8\pi \tilde{G} \tilde{\phi }_{*}^{2} \xi  \cdot  \left(- 480  \mu^2  -  2088 \mu \xi -  460 \mu  + 108   \xi^2   -  1055\xi   -  110  \right) \no \\
&+ \left( 8\pi \tilde{G}  \tilde{\phi }_{*}^2  \xi \right)^2  \cdot \left(-240 \mu^2   -  2088  \mu  \xi   - 230 \mu  \xi  -  5076  \xi ^2   -  1069  \xi   -  55  \right)  
\Bigg]\cdot \left( 384 \pi ^2 \left(1+8 \pi \xi \widetilde{G}\tilde{\phi }_{*}^{2} \right)^2 \cdot \Omega^2  \right) ^{-1} , \label{Appendix:beta-G-1L}
\end{align}
}%
{\small
\begin{align}
- \frac{1}{2} \left. \partial_t \xi \right|_{\text{1-loop}} & =  
  \Bigg\{
-3 \lambda  \left(10 \mu +28 \xi +5\right)   
 -8 \pi  \xi  \tilde{G} \cdot  \Big[3 \xi  \left(24 \mu^2 -77 \mu  +31 \right)  - \left(192 \mu^3 + 334 \mu^2 + 119\mu \right)
 +144 \xi ^2 \left(3 \mu+1\right)  \Big]  \no \\
 & + 64 \pi^2 \xi^2 \tilde{G}^2 \tilde{\phi }_{*}^{2}  \cdot \Big[3 \xi  \left(1056 \mu^2  +  528 \mu -  31\right) + \mu \left(384 \mu^2  +  758 \mu  +  283\right) + 9 \xi ^2 \left(276 \mu - 85 \right)-972 \xi ^3  \Big] \no \\
 & +  512 \pi ^3 \mu  \xi ^3 \tilde{G}^3 \tilde{\phi }_{*}^{4}  \cdot \Big[192 \mu^2+27 \xi \left(120 \mu + 47 \right)+  394 \mu+  972 \xi ^2  + 149  \Big]
 \Bigg\}  \cdot \left( 384 \pi^2  \Omega^3 \right)^{-1} , \label{Appendix:beta-xi-1L}
\end{align}
}%
%
%
{\small
\begin{align}
  &   \frac{1}{4} \left. \partial_t \lambda \right|_{\text{1-loop}}  =   \Bigg\{ 
 63 \lambda ^2
+ 16 \pi  \lambda  \tilde{G} \cdot   \Big[ 72 \mu^3+104 \mu^2 + 27 \xi ^2 \left(3 \mu+4\right) - 42 \xi  \left(\mu-1 \right)+  64 \mu  +15 \Big]  \no \\
  &    +96 \pi ^2 \xi  \tilde{G}^2   \cdot \Big[\xi ^2 \left(-468 \mu^2  +588 \mu  +6 \right)  + \xi \mu \left(1296 \mu^2+1658 \mu + 631\right)+4 \mu \left(72 \mu^3+104 \mu^2 + 64 \mu+15\right)+54 \xi ^3 \left(6\mu+1\right)\Big]\no \\
  &    +768 \pi ^3 \xi ^2 \tilde{G}^3 \tilde{\phi }_{*}^2 \cdot  \Big[  \xi ^2 \left(9144 \mu^2 + 4953 \mu + 6\right) + 2  \xi  \mu \left(1296 \mu^2 + 1826 \mu  +  547\right) +  4 \mu \left(72 \mu^3+104\mu^2+64 \mu+15\right)+54 \xi ^3 \left(6 \mu+1\right) \Big] \no \\
  &   +2048 \pi ^4 \mu \xi ^3 \tilde{G}^4 \tilde{\phi }_*^4 \cdot \Big[  9 \xi  \left(432\mu^2 + 632 \mu+173 \right)  +  4\left(72\mu^3 + 104 \mu^2 + 64\mu +15 \right)  +  81 \xi ^2 \left(360 \mu+167\right)+39366 \xi ^3 \Big]
   \no \\
   &  \hspace{2.3cm} \Bigg \} \cdot \left( 192 \pi ^2 \left( 1+8 \pi \xi \widetilde{G}\tilde{\phi }_{*}^{2} \right) \cdot \Omega^3  \right)^{-1}, \label{Appendix:beta-lambda-1L}
    \end{align}
    }%
along with 
\begin{align}
 \Omega \equiv 16 \pi  \mu  \xi  \tilde{G} \tilde{\phi }_*^{2}  + 8 \pi (9 \xi +1) \xi \tilde{G} \tilde{\phi }_*^{2}   + 2\mu +1, \; \; \; \; \mu \equiv \lambda \tilde{\phi }_*^{2}.
\end{align}
For completeness, here we present the equations for $\tilde{\phi }_*$ and $\tilde v$. To avoid presenting too many lengthy expressions it will be enough to present the corresponding expressions without gravitational corrections. They read as

\begin{align}
\left. \beta_{\tilde{\phi}_{*}^2 } \right|_{\widetilde{G}=0} =  - 2 \tilde{\phi}_{*}^2 + \frac{7}{32 \pi^2} \cdot  \frac{ 1 - 4 \mu}{ (1 + 2\mu)^3},
\label{beta-phi-Appendix}
\end{align}
\begin{align}
\left. \beta_{\tilde v} \right|_{\widetilde{G}=0} = - 4 \tilde v 
+  \frac{1}{192\pi^2} \cdot \frac{ 43 + 223 \mu + 481 \mu^2 + 288 \mu^3}{ (1 + 2\mu)^3}.
\label{beta-v-Appendix}
\end{align}

The expression for the running of the coupling $\xi$ in the limit $\tilde{u}, \tilde{\phi}_* \to 0$, and without invoking the 1--loop approximation assumes a relatively simple form and reads as,
\begin{align}
\frac{\partial_ t \xi}{\xi} = & \frac{4 \pi \lambda (252 \xi + 45) + \Big( 13824 \pi ^2 \xi ^3+8928 \pi ^2 \xi ^2  -84 \lambda  \xi -15 \lambda \Big) \cdot \tilde{G}   -  8\pi \Big(126 \xi ^4 - 204\xi^3  -  (997/2)   \xi ^2  + 55 \xi \Big)\cdot \tilde{G}^2  }  
{ 2304 \pi ^3 \xi + 384 \pi^2   \left(18\xi ^2  + 15\xi - 1 \right) \cdot \xi  \tilde{G}  -  16 \pi  \left(36 \xi ^2  +  30 \xi  -  1\right)\cdot  \xi  \tilde{G}^2 }. \label{xi-exact-Appendix}
\end{align}

\section{Initial conditions at the electroweak scale}\label{sec:EWTOINFL}
For completeness, here we report the 1-loop SM equations of Ref. \cite{DeSimone:2008ei} used to calculate the initial conditions for the SM couplings at the inflationary scale.
The couplings of interest are $\{\lambda, y_t, g_s, g_{EW}, g_{EM} \}$, corresponding to the quartic Higgs, the top-Yukawa, the strong, $SU(2)_L$ and $U(1)_Y$ gauge couplings respectively. The initial conditions we use at the top-quark mass scale are given by the following relations \cite{Buttazzo:2013uya}
\begin{align}
& \lambda =  0.12604, \; \; \;  y_t =0.9369 , \; \; \; g_s = 1.1666, \; \; \;  g_{\text{EW}} = 0.64779, \; \; \; g_{\text{EM}} =0.35830, \; \; \; \; \text{at $k = M_t = 173.34 GeV$} \label{measured-couplings-EW}.
\end{align}
Neglecting the contributions of the lighter quarks in the $MS$ scheme, the expressions presented in Ref. \cite{DeSimone:2008ei} read
\begin{align}
& \beta_\lambda^{\text{(SM)}} = \frac{1}{16 \pi^2}  \cdot  \left( 21 \lambda^2 + \lambda \cdot (  12y_t^2  - 3g_{\text{EM}}^2 - 9g_{\text{EW}}^2)   - 6 y_{t}^4  +  \frac{9}{8}g_{\text{EW}}^4 + \frac{3}{8}g_{\text{EM}}^4 + \frac{3}{4} g_{\text{EM}}^2 \cdot g_{\text{EW}}^2   \right), \no \\
& \beta_{y_t}^{\text{(SM)}} = \frac{y_t}{16 \pi^2} \cdot \left( \frac{9}{2} y_{t}^2 - 8 g_s^2 - \frac{17}{12}g_{\text{EM}}^2- \frac{9}{4}g_{\text{EW}}^2 \right),  \no \\
& \beta_{g_{\text{EM}}}^{\text{(SM)}}= \frac{1}{16 \pi^2} \cdot \frac{41}{6}   g_{\text{EM}}^3, \; \; \; \; \beta_{g_{\text{EW}}}^{\text{(SM)}} =  - \frac{1}{16 \pi^2} \cdot \frac{19}{6}  g_{\text{EW}}^3, \; \; \; \; \beta_{g_s}^{\text{(SM)}} = - \frac{7}{16 \pi^2} g_s^3. \label{betafuncs:SM} 
\end{align}
We have modified the coefficient of the $\lambda^2$ term in $\beta_\lambda^{\text{(SM)}}$ so that it matches the one derived here. To get the initial conditions at inflationary scales, we solve equations (\ref{betafuncs:SM}) for the initial conditions (\ref{measured-couplings-EW}) at $k = M_t$ up to the inflationary cut--off scale, following the spirit of \cite{Bezrukov:2008ej, DeSimone:2008ei, Bezrukov:2009db}. In doing so, we will neglect any gravitational corrections.
Using the measured values (\ref{measured-couplings-EW}) as initial conditions in the equations (\ref{betafuncs:SM}), and integrating them numerically up to $k = k_{\text{(inflation)}} \sim 10^{-4} \Mp$ (see Section \ref{sec:Q-dynamics-inflation}) we find
\begin{align}
& \lambda  = 6.461 \cdot 10^{-3}, \; \; \; y_t = 0.6292, \; \; \; g_{s} =0.7429, \; \; \;  g_{\text{EW}}  =  0.5903, \; \; \; g_{\text{EM}} = 0.3852   \label{Initial-conditions-SM-infl}.
\end{align}
Notice that for above initial conditions $\lambda$ crosses zero around the scale $\sim 10^{-3} \Mp$. The actual significance and implications for the (meta-) stability of the potential is a very delicate issue, also strongly depending on the initial conditions at the EW scale. This poses an important issue of study on its own, which we will not pursue here. In this work, it is enough for our purposes to assume that $\lambda$ remains positive at the inflationary scale, and we refer to \cite{Bezrukov:2014ipa, Bezrukov:2012sa, Bezrukov:2014ina, Degrassi:2012ry,EliasMiro:2011aa, Gies:2013fua, Eichhorn:2015kea,Litim:2015iea} for discussions. 

\acknowledgments
I am thankful to Astrid Eichhorn for communications with constructive criticisms and suggestions, as well as Ignacy Sawicki for illuminating discussions. I would further like to thank Kevin Falls, Mark Hindmarsh, Daniel Litim, Roberto Percacci and Christoph Rahmede for a constructive correspondence and feedback, as well as Ed Copeland and Tony Padilla for useful discussions.
This work has been supported by the Fundação para a Ciência e Tecnologia (FCT), through the Investigador research grant SFRH/BPD/95204/2013, as well as UID/FIS/04434/2013.

\bibliography{ASreferences.bib,Higgs_inflation.bib,UM.bib}
\end{document}